\documentclass[aps,prl,twocolumn,superscriptaddress]{revtex4-1}
\usepackage{graphicx}% Include figure files
\usepackage{dcolumn}% Align table columns on decimal point
\usepackage{bm}% bold math
\usepackage{color}
\usepackage{multirow}
\usepackage{amsmath}

\begin{document}
	\author{Yongqiang Li}
	\affiliation{Department of Physics, National University of Defense Technology, Changsha 410073, P. R. China}
	\affiliation{Department of Physics, Graduate School of China Academy of Engineering Physics, Beijing 100193, P. R. China}
	\author{Han Cai}
	\affiliation{Interdisciplanery Center for Quantum Information and State Key
Laboratory of Modern Optical Instrumentation, Zhejiang Province Key Laboratory of Quantum Technology and Device and Department of Physics, Zhejiang University, Hangzhou 310027, China}
	\author{Da-wei Wang}
	\affiliation{Interdisciplanery Center for Quantum Information and State Key Laboratory of Modern Optical Instrumentation, Zhejiang Province Key Laboratory of Quantum Technology and Device and Department of Physics, Zhejiang University, Hangzhou 310027, China}
	\author{Lin Li}
	\affiliation{MOE Key Laboratory of Fundamental Physical Quantities Measurement, Hubei Key Laboratory of Gravitation and Quantum Physics, PGMF and School of Physics, Huazhong University of Science and Technology, Wuhan 430074,  P. R. China}
	\author{Jianmin Yuan}
	\affiliation{Department of Physics, Graduate School of China Academy of Engineering Physics, Beijing 100193, P. R. China}
	\author{Weibin Li}
	\affiliation{School of Physics and Astronomy,  and Centre for the Mathematics and Theoretical Physics of Quantum Non-equilibrium Systems, The University of Nottingham, Nottingham NG7 2RD, United Kingdom}
	
	\title{Many-body chiral edge currents and sliding phases of atomic spinwaves in momentum-space lattice}
	\date{\today}
	\begin{abstract}
		Collective excitations (spinwaves) of long-lived atomic hyperfine states can be synthesized into a Bose-Hubbard model in momentum space. We explore many-body ground states and dynamics of a two-leg momentum-space lattice formed by two coupled hyperfine states. Essential ingredients of this setting are a staggered artificial magnetic field engineered by lasers that couple the spinwave states, and a state-dependent long-range interaction, which is induced by laser-dressing a hyperfine state to a Rydberg state. The Rydberg dressed two-body interaction gives rise to a state-dependent blockade in momentum space, and can amplify staggered flux induced anti-chiral edge currents in the many-body ground state in the presence of magnetic flux. When the Rydberg dressing is applied to both hyperfine states, exotic sliding insulating and superfluid/supersolid phases emerge. Due to the Rydberg dressed long-range interaction, spinwaves slide along a leg of the momentum-space lattice without costing energy.  Our study paves a route to the quantum simulation of topological phases and exotic dynamics with interacting spinwaves of atomic hyperfine states in momentum-space lattice.
	\end{abstract}
	\maketitle
	
	\textit {Introduction---}
	\begin{figure}[h!]
		\includegraphics[width=0.95\linewidth]{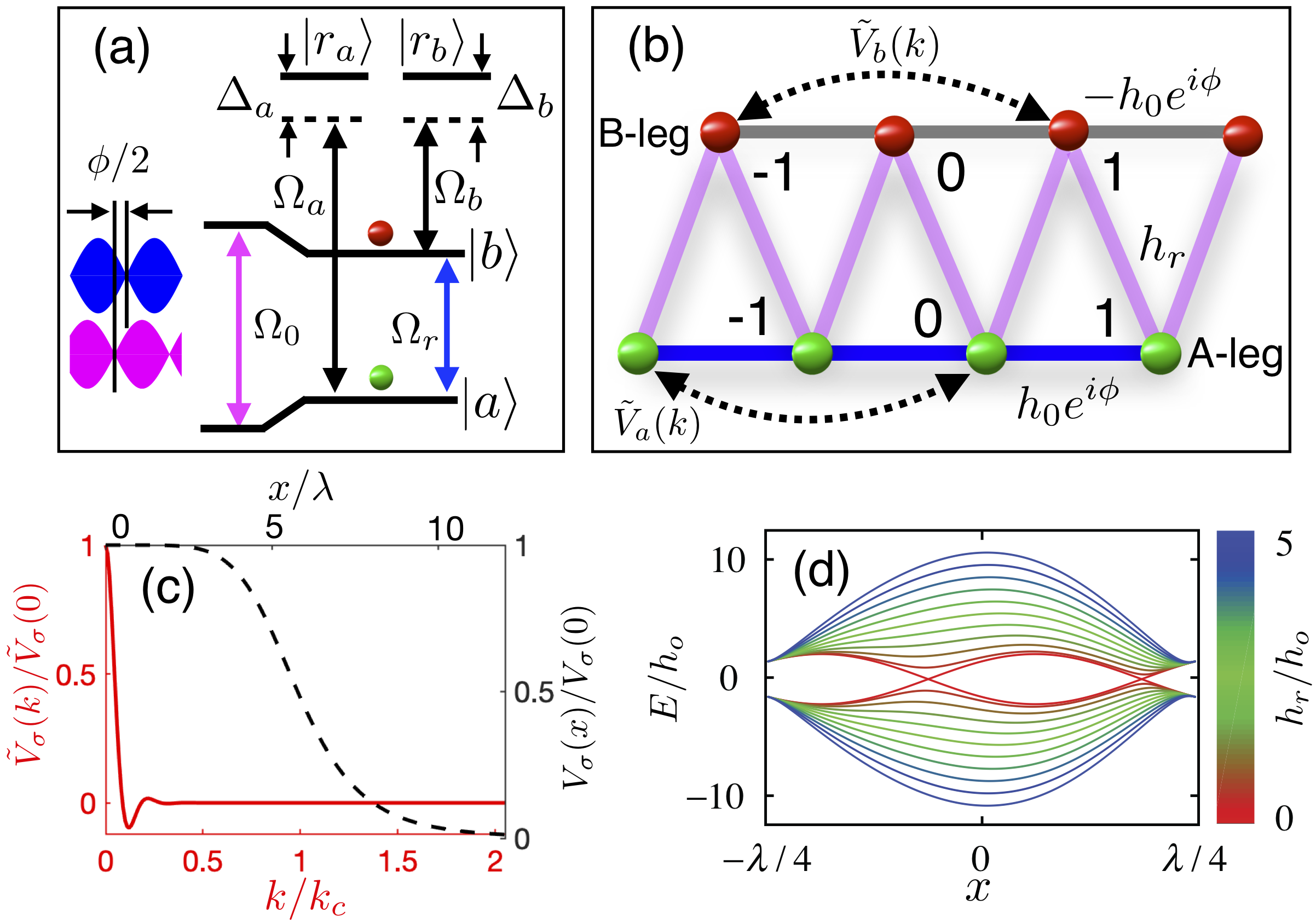}
		\vspace{-3mm}
		\caption{(Color online) \textbf{Momentum-space lattice}. (a) Level scheme. Collective excitations in states $|a\rangle$ and $|b\rangle$ are coupled resonantly by a detuned (magenta, Rabi frequency $\Omega_0$) and a resonant (blue, Rabi frequency $\Omega_r$) standing wave laser, forming a momentum-space lattice.  When weakly coupled to a Rydberg state (with Rabi frequency $\Omega_\sigma$ and detuing $\Delta_\sigma$), atoms in state $|\sigma\rangle (\sigma=a,b)$ experience an effective interaction $V_\sigma(x)$.  (b) Two-leg zig-zag lattice. The state $|a\rangle$ ($|b\rangle$) sits on the A-leg (B-leg) of the ladder. Hopping rate along the A-leg (B-leg) is $h_0e^{i\phi}$ ($-h_0e^{i\phi}$). The interleg hopping is determined by parameter $h_r$. The interactions between sites in the A-leg (B-leg) is determined by $\tilde{V}_a(k)$ [$\tilde{V}_b(k)$].  (c) Momentum dependent interaction $\tilde{V}_\sigma(k)/\tilde{V}_\sigma(0)$  (solid) and soft-core interaction ${V}_\sigma(x)/{V}_\sigma(0)$ (dashed). Using parameters $\lambda = 785$ nm and $r_c=4.5\,\mu$m, the interaction is important when $k\ll k_c$. (d) Band structure of the noninteracting ladder for different hopping amplitudes and flux $\phi=\pi/4$.}\label{fig:system}
	\end{figure}
	Chiral edge states have played an important role in understanding quantum Hall effects~\cite{Klitzing1980New, PhysRevLett.49.405,hasan_colloquium:_2010} in solid state materials~\cite{senthil_symmetry-protected_2015,sinova_spin_2015,hansson_quantum_2017}. Ultracold atoms exposed to artificial gauge fields provide an ideal platform to simulate chiral edge currents in and out of equilibrium. This is driven by the ability to precisely control and in-situ monitor~\cite{RevModPhys.80.885,lewenstein2012ultracold} internal and external degrees of freedom, and atom-atom interactions~\cite{Bromley2018Dynamics}. Chiral dynamics~\cite{Goldman2016Topological, RevModPhys.83.1523, goldman2014light, RevModPhys.91.015005} has been examined in the continuum space~\cite{Y2009Synthetic,Lin2010A},  ladders~\cite{Mancini2015Observation,Livi2016Synthetic,Atala2014Observation, Stuhl2015Visualizing,PhysRevLett.121.150403}, and optical lattices~\cite{Aidelsburger2011Experimental,PhysRevLett.108.225304,Aidelsburger2013Realization,Hirokazu2013Realizing,Gregor2014Experimental,Kennedy2015Observation,Fl2016Experimental,Asteria2018Measuring}. However, chiral states realized in the coordinate space require extremely low temperatures (typical in the order of a few kilo Hz) to protect the topological states from being destroyed by motional fluctuations~\cite{RevModPhys.91.015005}. Up to now, experimental observations of chiral phenomena in ultracold gases are largely at a single-particle level, due to unavoidable dissipations (e.g. spontaneous emission and heating)~\cite{Wall2016Synthetic,PhysRevLett.119.185701,Kolkowitz2016Spin,Tai2017Microscopy,Bromley2018Dynamics,Rey_2019}, while the realization of many-body chiral edge currents in ultracold atoms is still elusive.
	
	A key element to build many-body correlations is strong two-body interactions. Atoms excited to electronically high-lying (Rydberg) states provides an ideal platform due to their strong and long-range van der Waals interactions (e.g. a few MHz at several $\mu$m separation~\cite{RevModPhys.82.2313}). Benefiting directly from this interaction and configurable spatial arrangement~\cite{labuhn2016tunable,PhysRevLett.121.123603,omran2019generation}, Rydberg atoms have been used to emulate topological dynamics ~\cite{de2018experimental,celi2019emerging} in a time scale typically shorter than Rydberg lifetimes (typically $10\sim 100\mu$s). On the other hand, Rydberg dressed states, i.e. electronic ground states weakly mixed with Rydberg states~\cite{PhysRevA.65.041803,PhysRevLett.104.223002,Honer_10,Cinti_10,Li_12}, have longer coherence time (hundreds of ms) and relatively strong long-range interactions~\cite{Henkel_10}. Rydberg dressing has been demonstrated experimentally in magneto-optical traps~\cite{dress_trap}, optical tweezers~\cite{jau_entangling_2016} and lattices~\cite{zeiher2016many}. A remaining open question is to identify routes to create chiral states by making use of Rydberg dressed atoms.
	
	In this work, we propose a new lattice setup to explore chiral edge currents of an interacting many-body system via hybridizing long-lived atomic hyperfine states with electronically excited Rydberg states. By mapping to momentum space~\cite{wang_superradiance_2015}, collective excitations (atomic spinwaves) of ultracold gases form effective sites of an extended two-leg Bose-Hubbard model with competing laser-induced complex hopping and Rydberg dressed interactions. With this new setup, previously untouched many-body phases and correlated chiral dynamics can be realized in momentum-space lattice (MSL). Through dynamical mean-field calculations, we identify novel chiral edge currents in the many-body ground state. The strong dressed interaction suppresses edge currents, leading to a blockade in momentum space. By incorporating the Rydberg dressing to the two-legs, we find exotic translational-symmetry-broken sliding insulating and superfuid/supersolid phases in momentum-space lattice.
	
	\textit {The two-leg Bose-Hubbard model---}
	Collective atomic excitations stored in hyperfine states $|a\rangle$ and $|b\rangle$ [Fig.~\ref{fig:system}(a)] are created by an off-resonant and a resonant standing wave lasers along the $x$-axis [with wave vector $\mathbf{k} = k_c \hat{x}$ ($\hat{x}$ is the unit vector) and wave length $\lambda=2\pi/k_c$]. For small number of excitations, the spinwaves are described by free bosons~\cite{wang_superradiance_2015}. By projecting the spinwave to momentum space with a characteristic momentum $k_c$~\cite{wang_superradiance_2015}, we obtain a ladder of A-leg and B-leg for the $|a\rangle$ and $|b\rangle$ states [Fig.~\ref{fig:system}(b)], respectively. The $j$-th site of A-leg (B-leg) represents a collective state with wave vector $2jk_c$ [$(2j-1)k_c$].  When the standing wave lasers are phase mismatched [Fig.~\ref{fig:system}(a)], a synthetic magnetic field is generated~\cite{PhysRevLett.122.023601}.  This gives a nearest-neighbor hopping with complex amplitudes along and between the ladders, described by the Hamiltonian
	\begin{equation}
	H_c=-\sum_i[ h_r(b^{\dagger}_{i-1} a_i + a^\dagger_{i} b^{}_{i})
	+ h_oe^{i\phi}(a^\dagger_i a_{i+1} - b^{\dagger}_i b_{i+1} )+ \text{H.c.} \big] \nonumber
	\end{equation}
	where $\phi$, $h_o$ and $h_r$ are the flux, intra- and inter-leg hopping amplitudes.
	
	\begin{figure*}[t]
		\includegraphics*[width=0.96\linewidth]{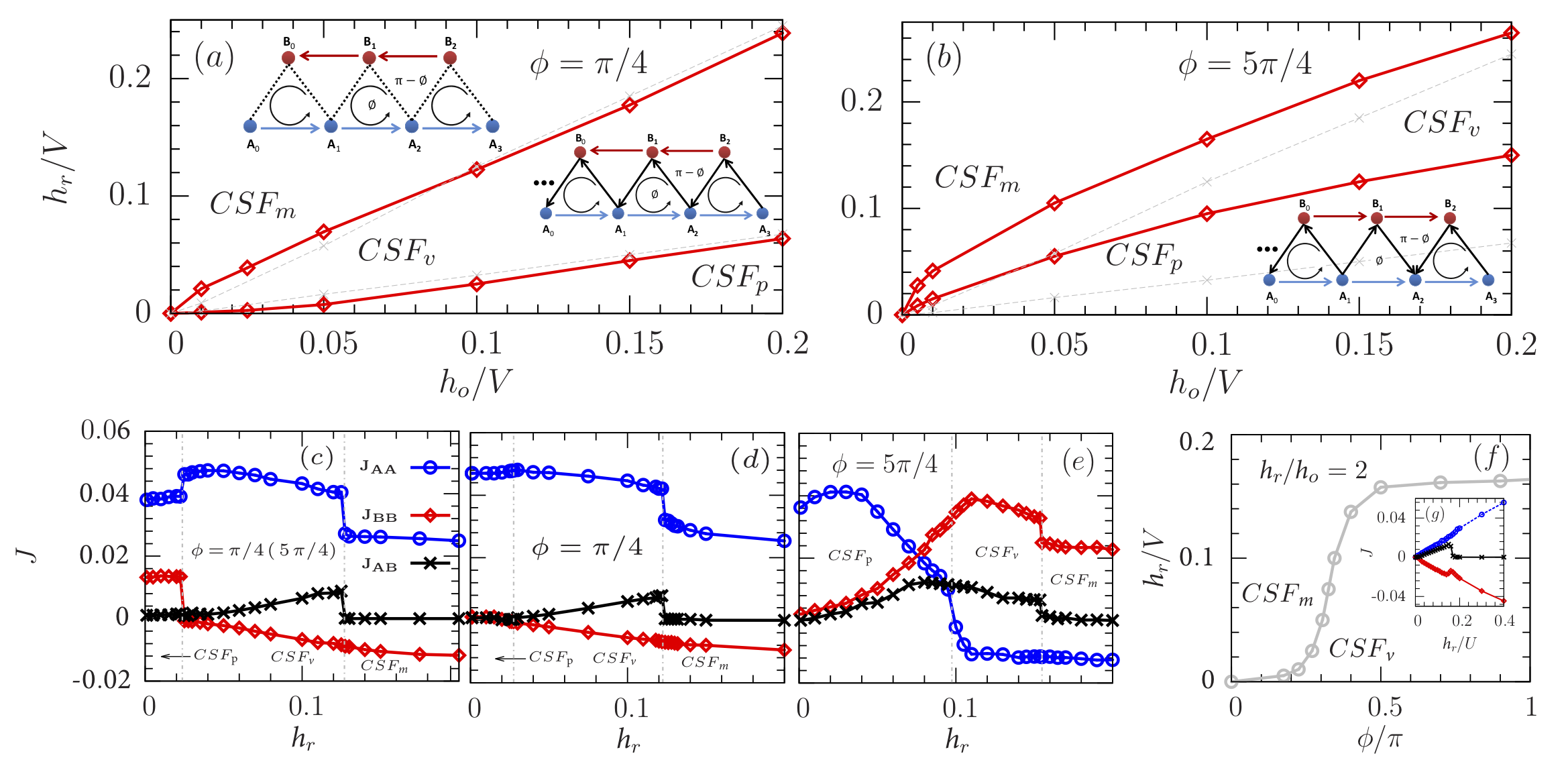}
		\caption{(Color online) {\textbf{Strongly correlated many-body ground states.} } (a)(b) Interaction effects on phase diagrams in momentum-space lattice in terms of hopping amplitude $h_r$ and $h_o$ for different flux. There are three quantum phases with different types of edge currents, including the CSF$_m$ and CSF$_v$ phases with chiral edge currents, and the CSF$_p$ phase with anti-chiral edge current. Here, the filling $N_{\rm tot}/L_{\rm lat}=0.125$ with $N_{\rm tot}$ being the total number of atoms in the lattice and $L_{\rm lat}$ the lattice size~\cite{explain_2}, and the dashed lines denote the non-interacting system. Interaction effects on phase transitions for a non-interacting (c) and interacting (d)(e) systems. Note that in the absence of two-body interactions, the currents for $\phi=5\pi/4$ can be obtained from (c) by using the chiral symmetry (i.e. swapping the state $|a\rangle$ and $|b\rangle$). The other parameters $h_o=0.1$ and $V=0$ (c); $h_o=0.1$ and $V=1$ (d); $h_o=0.1$ and $V=1$ (e). (f) Interplay between flux and interaction for a fixed hopping amplitude ratio. Inset: Interaction-induced CSF$_v$-CSF$_m$ phase transition for a fixed flux $\phi=0.6\pi$ (g). Periodic boundary condition is used in the calculation.}
		\label{phase_diagram}
	\end{figure*}
	In our setting, state $|a \rangle$ ($|b \rangle$) is coupled to a Rydberg state $|R_a\rangle$ ($|R_b\rangle$) by an off-resonant laser. This induces a state-dependent two-body soft-core shape interaction $V_{\sigma}(x)\equiv C_{\sigma} / (r_{\sigma}^6 +|x|^6)$, where $C_{\sigma}$ and $r_{\sigma}$ ($\sigma = a,b$) are the dispersion coefficient and characteristic distance of the interaction potential, respectively. Their values can be engineered by tuning parameters of the dressing laser ~\cite{Henkel_10,Honer_10,Cinti_10,Li_12}.
	The Hamiltonian for the interactions of the dressed state is,
	\begin{eqnarray}\label{eq:interaction}
	H_d= \sum_{p, i, l} \tilde{V}_a(p) a^\dagger_{i+p} a^\dagger_{l-p} a_{i}  a_{l} + \sum_{p, i, l} \tilde{V}_b(p) b^\dagger_{i+p} b^\dagger_{l-p} b_{i}  b_{l}, \nonumber
	\end{eqnarray}
	where $\tilde{V}_{\sigma}(k) = \sum_x {\rm exp}^{-i \pi k\cdot x}V_{\sigma}(x)$ is the Fourier transformation of the corresponding interaction. $\tilde{V}_{\sigma}(k)$ decays rapidly with increasing $k$  as typically $r_c\gg \lambda$ [Fig.~\ref{fig:system}(c)]. Finally the two-leg Bose-Hubbard Hamiltonian is given by $H=H_c + H_d -\sum_{i,\sigma} \mu_{\sigma} n_{i\sigma}$ with $n_{i\sigma}=\sigma^\dagger_{i}\sigma_i$ and $\mu_{\sigma}$ to be the atomic density at site (momentum) $i$ and chemical potential in state $|\sigma\rangle$. Details of the Hamiltonian can be found in the Supplemental Material (\textbf{SM}). In the following, we choose $V\equiv \tilde {V}_{b}(0)$ as the unit of the energy.
	
	\textit {Symmetry and ground state phases in the interaction free case---}
	The coupling Hamiltonian $H_c$ possesses [$\mathcal{\tau}H(\phi) \mathcal{\tau}^{-1} = H(-\phi)$], [$\mathcal{C}H(\phi)\mathcal{C}^{-1} = H(\pi+\phi)$], and [$\mathcal{T}H(\phi) \mathcal{T}^{-1} = H(2\pi+\phi)$], where $\mathcal{\tau}$, $\mathcal{C}$ and $\mathcal{T}$ are the time-reversal, chiral and translational symmetry operators. The chiral symmetry shows that currents after swapping the two states will remain the same if the flux is shifted simultaneously by $\pi$, i.e. $\phi\to \phi+\pi$. The system does not preserve the time-reversal symmetry in general except when $\phi=\pi/2$, where the ground state energy exhibits double degeneracy~\cite{PhysRevLett.122.023601}, already leading to rich chiral phases (see examples in \textbf{SM}).
	
	Solving $H_c$ exactly we obtain three types of band minima, i.e. a single minimum at real-space coordinate $x=0$ or $\neq0$, and two minima, as show in Fig.~\ref{fig:system}(d), analogous to spin-orbit coupling setup in real space~\cite{goldman2014light} (the states $|a\rangle$ and $|b\rangle$ treated as spin $|\uparrow\rangle$ and $|\downarrow\rangle$, respectively). The corresponding chiral phases are characterized by current $J_{\sigma\sigma'}\equiv-2{\rm Im}(h_{i,j}\langle \sigma^\dagger_i \sigma'_{j}\rangle)$ for the bond $j\rightarrow i$ and the leg $\sigma\rightarrow\sigma^\prime$, similar to the definition in real-space lattices~\cite{vasic2015chiral, plekhanov2018emergent}. As shown in Fig.~\ref{phase_diagram}, the ground state prefers a number of chiral superfluid phases ($\text{CSF}$s). When the inter-leg coupling $h_r$ is strong, currents along the ladders (edges) have opposite directions, i.e. $J_{AA} \times J_{BB} < 0$ and $J_{AB}=0$ [Fig.~\ref{phase_diagram}(c)], denoted by $\text{CSF}_m$ phase (condensed into band minimum at $x=0$). When $h_r$ and $h_o$ are comparable, we have a different chiral superfluid phase ($\text{CSF}_v$ condensed into band minimum at $x\neq0$) where currents on the ladders and rungs satisfy $J_{AA}\times J_{BB}<0$, and $J_{AB} \neq 0$. The $\text{CSF}_m$ and $\text{CSF}_v$ phases are analogues of the Meissner and vortex phases of real-space ladder systems~\cite{Mancini2015Observation,Livi2016Synthetic,Atala2014Observation, Stuhl2015Visualizing,Orignac2000Meissner, PhysRevLett.108.133001,PhysRevLett.111.150601, PhysRevLett.112.043001,PhysRevB.91.054520,Tokuno_2014,PhysRevB.91.140406,PhysRevA.94.063628, Anisimovas2016Semi, PhysRevX.7.021033, PhysRevA.95.023607,sundar2018synthetic,Barbiero_ladder_2019}.
	
When $h_r$ is further decreased, two band minima become degenerate [see Fig.~\ref{fig:system}(d)].  At the minima, currents on both ladders flow in the same direction, i.e. $J_{AA}\times J_{BB}>0$. This leads to a particularly interesting {\it anti-chiral edge current} (denoted by CSF$_p$) in the system. Compared to real-space lattices~\cite{Atala2014Observation,Anisimovas2016Semi}, a fundamental difference of our scheme is the staggered magnetic flux, {\it i.e.} the flux of a plateau in the zigzag ladder is $\phi$ and its neighbor $\pi-\phi$. The system then hosts asymmetric band structures in the regime $h_o/h_r \gg 1$, as shown in Fig.~\ref{fig:system}(d), with atoms in A- (B-) leg favoring the left (right) minimum [Fig.~\ref{momentum}(b)]. As a result, the staggered flux induced asymmetric condensation of particles in each ladder changes directions of edge currents (a perturbative derivation for $h_o/h_r \gg 1$ can be found in \textbf{SM}). As far as we know, this new quantum phase has not been studied in real-space lattice. Since the CSF$_p$ phase emerges in the limit $h_o/h_r \gg 1$, one could create this phase through an adiabatic manner. Starting with large $h_r$ and $h_o$, one can adiabatically reduce (increase) $h_r$ ($h_o$) to enter the CSF$_p$ phase region while keeping $h_o$ ($h_r$) fixed. The  anti-chiral edge current will be induced and saturate to a finite value even though $h_r/h_o\to 0$.
%Details of the condensation distribution and currents of the CSF$_p$ phase are shown in \textbf{SM}. ,%,  which is analogue to the ground state persistent current in normal metal rings threading an external magnetic flux~\cite{London_1937, PhysRevLett.7.46, PhysRev.137.A787,BUTTIKER1983365,Bleszynski2009Persistent},where the condensing positions for the states $|a\rangle$ and $|b\rangle$ are separated in real space, as shown in Fig.~S2 in appendix
	
	\textit {Stable anti-chiral currents and excitation blockade by single Rydberg dressing. ---}
	In the single Rydberg dressing the B-leg is coupled to a Rydberg state (i.e. $V_a(x)= 0$ and $V_b(x)\neq 0$). We employ a bosonic dynamical mean-field calculation that captures both quantum fluctuations and strong correlations in a unified framework~\cite{Li2011,vasic2015chiral, plekhanov2018emergent} (see \textbf{SM} for details). The reliability of this approach has been confirmed by a comparison with an unbiased quantum Monte Carlo simulation~\cite{PhysRevLett.105.096402}.
	
	Intuitively, one would expect currents are suppressed by the two-body interaction. On the contrary, the CSFs phases are stable against the dressed interaction [see Fig.~\ref{phase_diagram}(a) and (b)]. Actually, the two-body interaction breaks the chiral symmetry [$\mathcal{C}H(\phi)\mathcal{C}^{-1} = H(\pi+\phi)$]. The broken symmetry is most apparent in the CSF$_p$ phase, where the phase region shrinks when $\phi=\pi/4$ [Fig.~\ref{phase_diagram}(a)] but expands when $\phi=5\pi/4$ [Fig.~\ref{phase_diagram}(b)]. When $\phi=5\pi/4$, the two-body interaction reduces the energy separation between the two legs. As a result, both currents $J_{BB}$ and $J_{AB}$ are increased in the intermediate hopping regime. Here we observe a discontinuous phase transition in Fig.~\ref{phase_diagram}(c), and signatures of continues CSF$_p$-CSF$_m$ phase transition by increasing $h_r$  [Fig.~\ref{phase_diagram}(d)(e)]. Furthermore, one can drive transitions between the CSF phases by varying the flux $\phi$. One example can be found in Fig.~\ref{phase_diagram}(f), where the Meissner phase is driven to the vortex phase by changing the flux.
	
\begin{figure}[t]
		\includegraphics*[width=0.95\linewidth]{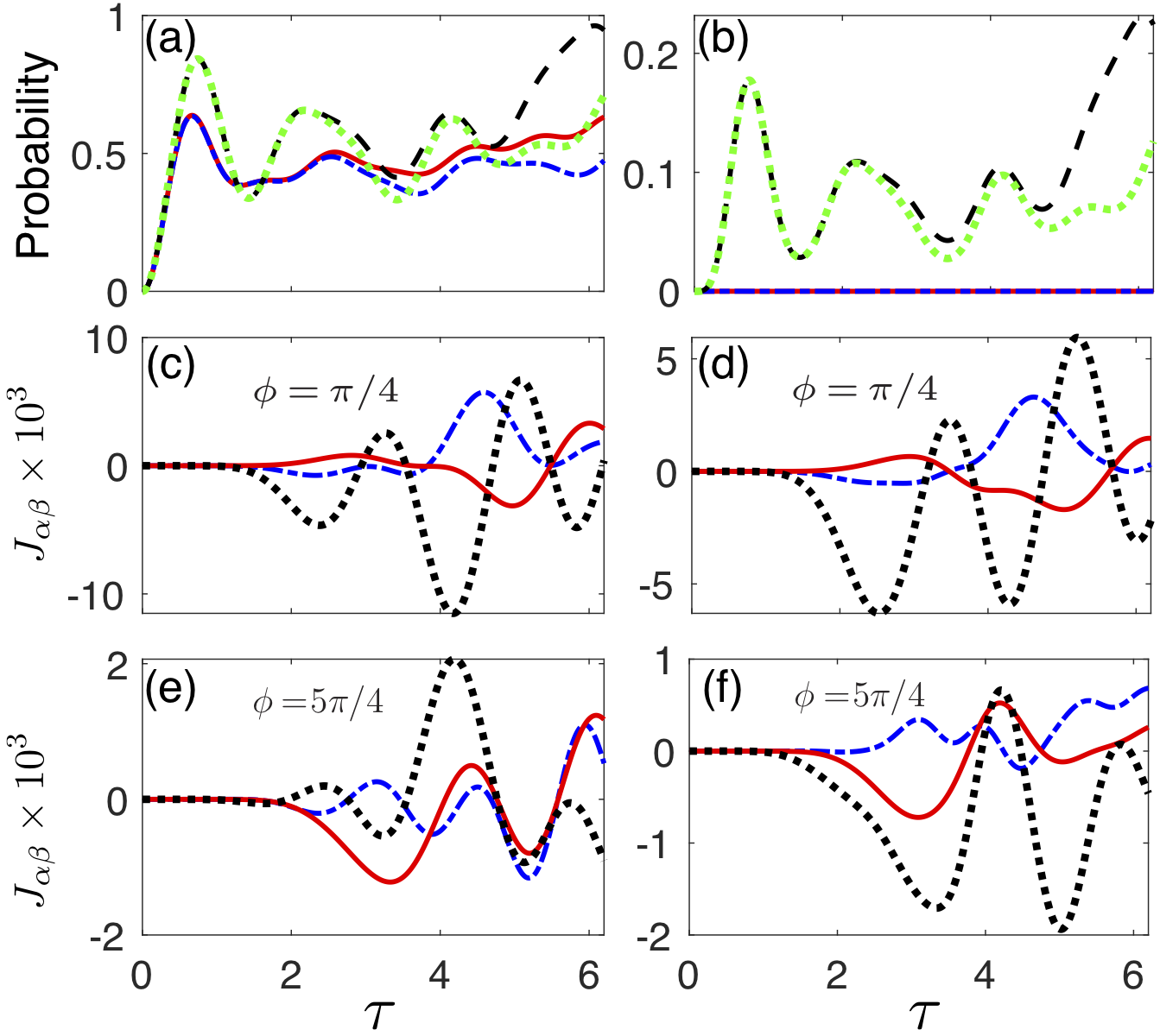}
		\caption{(Color online) {\textbf{Dynamics of excitations and currents}.} (a) Total excitation and (b) double occupation probability of the B-leg. Dotted and dashed curves correspond to non-interacting situations and dot-dashed and solid with two-body interactions. The flux is $\phi = \pi/4$ (solid and dashed) and $\phi=5\pi/4$ (dot-dashed and dotted). Time evolution of currents in the absence (c)(e) and presence (d)(f) of two-body interactions, with currents along the A-leg $J_{AA}$ (dotted), B-leg $J_{BB}$ (solid), and between the two legs $J_{AB}$ (dashed). Other parameters are $h_o=h_r=0.02$ and $V=0$ (c)(e), and $h_o/V=h_r/V=0.02$ (d)(f). We have defined $\tau \equiv h_0t$.
		}
		\label{fig:blockade}
	\end{figure}

	The two-body interaction plays a vital role in the dynamics. This is illustrated with a situation where initially two excitations at the first site of the A-leg (i.e. $n_1^{(a)}=2$) are prepared (see \textbf{SM} for details). When the atomic interaction is negligible, a large fraction of the excitation is transferred to the B-leg, as shown in Fig.~\ref{fig:blockade}(a). Turning on the interaction, the excitation probability in the B-leg is reduced apparently. A closer examination shows that the double occupation probability $P_2=\sum_j|\langle \psi(t)|(b_j^{\dagger})^2/\sqrt{2}|\psi(0)\rangle|^2$ in the B-leg is significantly suppressed by the strong interaction.  $|\psi(t)\rangle=\exp(-iHt)|\psi(t=0)\rangle$ is the many-body state at time $t$. This excitation suppression could be regarded as an interaction blockade in momentum space, analogue to the Rydberg blockade effect in real space~\cite{urban_observation_2009,gaetan_observation_2009}.
	\begin{figure}[t]
		\includegraphics*[width=0.98\linewidth]{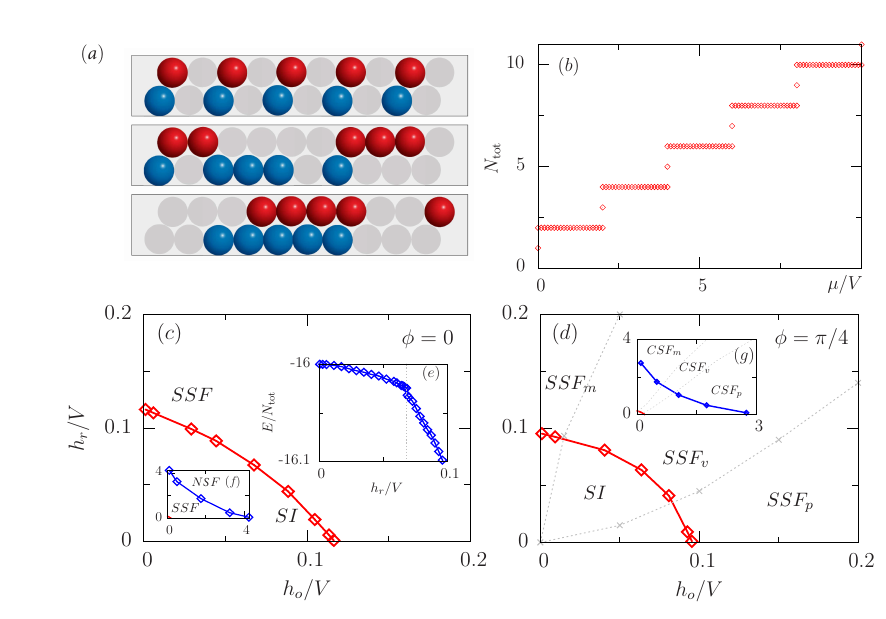}
		\caption{(Color online) {\textbf{Emergent sliding phases in the double Rydberg dressing regime.}} (a) Cartoon picture of sliding insulator with random density distribution with degenerate energy. (b) Devil's staircase structure of the system in the atomic limit, where $N_{\rm tot}$ denotes the total number of particles. (c) Phase diagram of the system with flux $\phi=0$ and particle number $N_{\rm tot}=32$, obtained via bosonic mean-field theory. The system supports the sliding insulator (SI) and sliding superfluid/supersolid (SSF). (d) Phase diagram with flux $\phi=\pi/4$ and particle number $N_{\rm tot}=32$. Both SI and SSF phases generate edge currents in the presence of flux. Inset (e): Averaged energy $E/N_{\rm tot}$ of the system as function of hopping $h_r=h_o$ in the strongly interacting regime. We observe a phase transition from sliding insulator to sliding superlfuid. Inset (f)(g): Zoom of the main figure, where normal superfluid phase NSF appears in the weakly interacting regime (CSFs appearing in the presence of flux).}
		\label{fig:sliding}
	\end{figure}
	
	The dynamically generated currents show features of the ground state phases, though their strengths are suppressed by the interaction. For example, when the parameters are in the CSF$_v$ phase, for example, the currents counter-propagate on the two legs at the beginning of the evolution [Fig.~\ref{fig:blockade}(c)-(e)]. Surprisingly, the currents for $\phi=5\pi/4$ along the legs become co-propagating when two-body interactions are important [Fig.~\ref{fig:blockade}(f)]. This behavior is similar to the currents found in the CSF$_p$ phase.
	
	\textit {Sliding phases when both chains are laser dressed to Rydberg states.---}
	A double Rydberg dressing is realized when the two legs are coupled to two different Rydberg states. For energetically close Rydberg states, this leads to similar dressed potentials for the two legs. Without loss of generality, we focus on the case with $\tilde{V}_a(k)=\tilde{V}_b(k)$ in the following.
	
Without hopping  ($h_r=h_o=0$),  the ground state is an unusual insulating state with a high  degree of degeneracy $g=C(N_a+L_{a}-1,N_a)\times C(N_b+L_{a}-1,N_b)$, where $C(j,k)= j!/[(j-k)!k!]$ is the binomial coefficient, $N_a$ and $N_b$ denote the number of particles, and $L_a$ and $L_b$ the lattice number in A- and B-legs, respectively. Its energy $E=\tilde{V}_a(0)N_a(N_a-1)/2+\tilde{V}_b(0)N_b(N_b-1)/2-\mu_a N_a - \mu_b N_b$ [$\tilde {V}_\sigma(0)$ dominates when $R_c\gg \lambda$] depends on the total number of particles but not their distributions in the momentum-space ladder, i.e. changing their locations in the corresponding leg costs no energy. In Fig.~\ref{fig:sliding}(a), we illustrate this phase with an example of five bosons along each leg. We attribute this phase as a sliding insulating phase (SI), analogous to long-sought sliding superfluid phase~\cite{PhysRevB.31.4516, PhysRevB.33.4767, PhysRevLett.80.4345, PhysRevLett.83.2745, PhysRevLett.105.085302, Linxiao2018Observation}. The underlying physics of the emergent SI studied here is a result of momentum-space Rydberg coupling (all sites equally coupled) from the long-range real-space Rydberg interaction. Varying the chemical potential, the ground state forms Devil's staircase structures with non-integer filling [Fig.~\ref{fig:sliding}(b)], which breaks the lattice translational symmetry.
	
	Turning on the hopping, the sliding insulating phase is stable in the low hopping regime ($h_o\,, h_r \ll V$), as shown in the phase diagram Fig.~\ref{fig:sliding}(c) and (d) (detailed discussions for finite hopping shown in \textbf{SM}). For larger hopping amplitudes, density fluctuations are stronger and a sliding superfluity phase (SSF) appears.  The SI-SSF phase transition can be determined by examining the ground state energy [Fig.~\ref{fig:sliding}(e)]. Note that the SSF breaks both lattice translational and gauge symmetries, and therefore can be considered as a sliding supersolid phase. In the weakly interacting regime ($h_o\,, h_r \gg V$), kinetic energy dominates and normal superfluid (NSF) appears [Fig.~\ref{fig:sliding}(f)(g)]. In the presence of external magnetic flux, the chiral symmetry is broken, i.e. nonzero local currents are found in the sliding phases. In the weakly interacting limit, normal chiral superfluid phases emerge in the ground state.
	
	\textit{Conclusion---} A momentum-space lattice model suitable for investigating topological physics and correlated many-body dynamics is proposed. Using single and double Rydberg dressing,  the ground state exhibits a series of exotic phases, including anti-chiral edge currents, sliding insulating and superfluid states. Strongly correlated dynamics, such as the state dependent excitation blockade, is found due to the competition between the chirality and strong Rydberg interactions. Compared to schemes based on superradiant Dicke states characterized by steady states~\cite{wang_superradiance_2015,PhysRevLett.120.193601,PhysRevLett.122.023601}, our setting permits to explore correlated chiral phenomena both in and out of equilibrium coherently. Compared to real-space magnetic ladder system, our system is essentially a spin-orbit coupling setup in momentum space, and supports three distinctive chiral broken phases.
	
	Our study paves new routes towards the study of chirality with interacting spinwaves in higher dimensions and external driving. In frustrated lattices (e.g. honeycomb lattices), emergent quantum topological dynamics can be investigated in a Hamiltonian system, e.g., by quenching the system from a trivial to topological state. When dissipation is introduced, this opens opportunities to uncover stability of edge modes, as well as to explore competing dynamics between atomic interactions and dissipation in an open quantum system (example of quantum Zeno effect is discussed  in \textbf{SM}.).
	
	\section{Acknowledgements}
	Authors would like to thank helpful discussions with Jing Zhang, Yinghai Wu, XiongJun Liu, and Tao Shi. This work is supported by the National Natural Science Foundation of China under Grants No. 11304386 and No. 11774428 (Y. L.), by the UKIERI-UGC Thematic Partnership No. IND/CONT/G/16-17/73, EPSRC Grant No. EP/M014266/1 and EP/R04340X/1 (W. L.), and by National Natural Science Foundation of China (No. 11874322), the National Key Research and Development Program of China (Grant No. 2018YFA0307200) (D. W.). The work was carried out at National Supercomputer Center in Tianjin, and the calculations were performed on TianHe-1A.

\clearpage
\begin{widetext}
\begin{center}
{\Huge \bf Supplementary Material}
\end{center}
\renewcommand{\theequation}{S\arabic{equation}}
\renewcommand{\thesection}{S-\arabic{section}}
\renewcommand{\thefigure}{S\arabic{figure}}
\renewcommand{\bibnumfmt}[1]{[S#1]}
\renewcommand{\citenumfont}[1]{S#1}
\setcounter{equation}{0}
\setcounter{figure}{0}

\section{Extended Bose-Hubbard model in momentum space}
\subsection{The Hamiltonian in momentum space}
Here we consider N three-level atoms, i.e. ground state $g(\bf{r}_i)$, another ground  state $a(\bf{r}_i)$ and Rydberg dressed state $b(\bf{r}_i)$, where $\bf{r}_i$ is the  position of the $i$th atom with random distribution. The atoms are initially prepared in the ground state $|G\rangle\equiv|g_1...g_N\rangle$. A standing wave laser couples the atomic $a$ and $b$ states with vectors $\bf{k}_1=-\bf{k}_2=\bf{k}_c$. In the rotating wave approximation, the Hamiltonian reads

\begin{eqnarray}
H = \big(-\sum_j h_r(e^{i\bf{k}_1\cdot\bf{r}} + e^{i\bf{k}_2\cdot\bf{r}})|b_j\rangle\langle a_j|+{\rm H.c.}\big)+  \sum^N_{i\neq j}\frac{C_b}{r_b^6 +|\mathbf{r}|^6}  |...b_i..b_j...\rangle\langle ...b_i...b_j...|,
\end{eqnarray}
where $C_b$ and $r_b$ is the dispersion coefficient and characteristic distance of the soft-core shape interaction, respectively~\cite{Henkel_10,Honer_10,Cinti_10,Li_12}. Collective atomic excitation operators in momentum space are introduced as
 \begin{eqnarray}
a^\dagger_l \equiv \frac{1}{\sqrt{N}}\sum^N_{i=1} e^{i2lk_cr_i} |...a_i...\rangle\langle G|,
\end{eqnarray}
 \begin{eqnarray}
 b^{\dagger}_l \equiv \frac{1}{\sqrt {N}}\sum^N_{i=1} e^{i(2l-1)k_cr_i}  |...b_i...\rangle\langle G|,
\end{eqnarray}

We transform the Hamiltonian from position space to momentum space via

 \begin{eqnarray}
 |...a_i...\rangle\langle G| \equiv \frac{1}{\sqrt{N}}\sum_le^{-i2lk_cr_i}a^\dagger_l,
 \end{eqnarray}

 \begin{eqnarray}
 |...b_i...\rangle\langle G| \equiv \frac{1}{\sqrt{N}}\sum_le^{-i(2l-1)k_cr_i}b^\dagger_l.
 \end{eqnarray}

The total Hamiltonian in momentum space can be written as
\begin{eqnarray}
H = &-& \sum_{i}h_r(b^{\dagger}_{i-1} a_i + a^\dagger_{i} b_{i}) + {\rm H.c.} \nonumber \\
&+& \sum_{i_1, i_2, i_3, i_4} \tilde{V}_b(\Delta i) b^\dagger_{i_1} b_{i_2} b^\dagger_{i_3} b_{i_4}\delta_{i_1 - i_2 + i_3 - i_4},
\end{eqnarray}
where $\tilde{V}_b(\Delta i)\equiv \tilde{V}_b(i_1 - i_2) = \sum_r e^{-\pi i(i_1 - i_2)kr}V_b(r)$, and the transformation is valid for many excitations if the excitation number is much less than the atom number \cite{wang_superradiance_2015}.

Here, if we switch on another far-tuned standing wave lasers and couple the $|a \rangle$ state to another Rydberg state, then an extra interaction term $-\sum_j2h_o\cos(2k_cr_j+\phi)(|a_j\rangle\langle a_j|-|b_j\rangle\langle b_j|)$ appears due to the AC Stark shifts \cite{PhysRevLett.122.023601}. The total Hamiltonian is given by:
\begin{eqnarray}\label{eff_Ham}
H =&-& \sum_{i}h_r( b^{\dagger}_{i-1} a_i + a^\dagger_{i} b_{i}) + {\rm H.c.} - \sum_{i} e^{i\phi}h_o (a^\dagger_i a_{i+1} - b^{\dagger}_i b_{i+1} ) + {\rm H.c.}  \nonumber \\
&+& \sum_{p, i, l} \tilde{V}_a(p) a^\dagger_{i+p} a^\dagger_{l-p} a_{i}  a_{l}+ \sum_{p, i, l} \tilde{V}_b(p) b^\dagger_{i+p} b^\dagger_{l-p} b_{i}  b_{l}.
\end{eqnarray}

\section{bosonic dynamical mean-field theory}
To investigate ground states of bosonic gases loaded into momentum-space lattices, described by Eq.~(\ref{eff_Ham}), we establish a bosonic version of dynamical mean-field theory (BDMFT) on the ladder system with $z=4$, where $z$ is the number of neighbors connected by hopping terms. As in fermionic dynamical mean field theory, the main idea of the BDMFT approach is to map the quantum lattice problem with many degrees of freedom onto a single site - "impurity site" - coupled self-consistently to a noninteracting bath~\cite{Appen_georges96}. The dynamics at the impurity site can thus be thought of as the interaction (hybridization) of this site with the bath. Note here that this method is exact for infinite dimensions, and is a reasonable approximation for neighbors $z\ge4$. In the noninteracting limit, the problem is trivially solvable in all dimensions, all correlation functions factorize and the method becomes exactly~\cite{Appen_Byczuk_2008}.

\subsection{BDMFT equations}
In deriving the effective action, we consider the limit of a high but finite dimensional optical lattice, and use the cavity method~\cite{Appen_georges96} to derive self-consistency equations within BDMFT. In the following, we use the notation $h_{ij}$ for the hopping amplitude between sites $i$ and $j$, and define creation field operator $d^\dagger$ for the state $|\sigma\rangle$ [$\sigma=a(b)$] to shorten Ham.~(\ref{eff_Ham}). And then the effective action of the impurity site up to subleading order in $1/z$ is then expressed in the standard way~\cite{Appen_georges96, Appen_Byczuk_2008}, which is described by:
\begin{eqnarray}\label{eff_action}
S^{(0)}_{\rm imp} &=& -\int_0^\beta \hspace{-0.2cm} d \tau d\tau' \sum_{\sigma\sigma'} \Bigg( \hspace{-0.1cm} \begin{array}{c} d^{(0)*} (\tau)\quad d^{(0)} (\tau) \end{array}\hspace{-0.1cm} \Bigg)^{\hspace{0.1cm}} \boldsymbol{\mathcal{G}}^{(0)-1}(\tau-\tau') \Bigg(\begin{array}{c} \hspace{-0.1cm} d^{(0)} (\tau')\\ d^{(0)*} (\tau') \end{array} \hspace{-0.1cm}\Bigg) 	 \\
&+& \int_0^\beta d\tau \sum_{j,p}\tilde{V}_d(p) d^{(i+p)*}(\tau)  d^{(j-p)*}(\tau) d^{(i)}(\tau) d^{(j)}(\tau)  \nonumber
,\label{action}
\end{eqnarray}
with Weiss Green's function
\begin{eqnarray}
&&\hspace{-5mm}\boldsymbol{\mathcal{G}}^{(0)-1}(\tau-\tau') \equiv - \\
&&\hspace{-5mm} \left(\begin{array}{cc} \hspace{-0.1cm}
				(\partial_{\tau'}-\mu)\delta+
\hspace{-0.25cm} \sum \limits_{\langle 0i\rangle,\langle 0j\rangle} \hspace{-0.25cm}
h_{ij}^2G_{ij}^1 (\tau, \tau')
			&   \hspace{-0.25cm} \sum \limits_{\langle
0i\rangle,\langle 0j\rangle} \hspace{-0.25cm} h_{ij}^2   G^2_{ij}(\tau, \tau') \\
			  \hspace{-0.25cm} \sum \limits_{\langle
0i\rangle,\langle 0j\rangle} \hspace{-0.25cm} h_{ij}^2 {G^2_{ij}}^*(\tau', \tau)
	 & (-\partial_{\tau'}-\mu_\sigma)\delta+   \hspace{-0.25cm}
\sum \limits_{\langle 0i\rangle,\langle 0j\rangle} \hspace{-0.25cm} h_{ij}^2  G_{ij}^1
(\tau', \tau)
			 \hspace{-0.1cm} \end{array}\right) \hspace{-0.15cm},	\nonumber
\end{eqnarray}
and superfluid order parameter
\begin{equation}
\Phi^{}_{i}(\tau) \equiv \langle d_{i} (\tau) \rangle_0.
\end{equation}
Here, we have defined the the diagonal and off-diagonal parts of the connected Green's functions
\begin{eqnarray}
\hspace{-0.5cm}G_{ij}^1 (\tau, \tau')\hspace{-0.2cm}&\ \equiv\ &\hspace{-0.2cm}- \langle d_{i} (\tau) d_{j}^* (\tau')
\rangle_0 + \Phi_{i} (\tau) \Phi_{j}^* (\tau'), \\
\hspace{-0.5cm}G_{ij}^2 (\tau, \tau')\hspace{-0.2cm}&\ \equiv\ &\hspace{-0.2cm}- \langle d_{i} (\tau) d_{j} (\tau')
\rangle_0 + \Phi_{i} (\tau) \Phi_{j} (\tau'),
\end{eqnarray}
 where $\langle \ldots \rangle_0$ denotes the expectation value in the cavity system (without the impurity site).

To find a solver for the effective action, we return back to the Hamiltonian representation and find that the local Hamiltonian is given by a bosonic Anderson impurity model.
\begin{eqnarray}
\hat{H}^{(0)}_A &=& - \sum t \Big(\Phi^{(0)*} \hat{d}^{(0)} + {\rm H.c.} \Big) + \sum_{i,j,p}\tilde{V}_d(p) \hat{d}^\dagger_{j+p} \hat{d}^{\dagger}_{i-p} \hat{d}_j \hat{d}_{i} + \sum_{l}  \epsilon_l \hat{a}^\dagger_l\hat{a}_l + \sum_{l} \Big( V_{l} \hat{a}_l\hat{d}^{\dagger(0)} + W_{\sigma,l} \hat{a}_l\hat{d}^{(0)} + {\rm H.c.} \Big),
\end{eqnarray}
where the chemical potential and interaction term are directly inherited from the Hubbard Hamiltonian. The bath of condensed bosons is represented by the Gutzwiller term with superfluid order parameters $\Phi^{(0)}$. The bath of normal bosons is described by a finite number of orbitals with creation operators $\hat{a}^\dagger_l$ and energies $\epsilon_l$, where these orbitals are coupled to the impurity via normal-hopping amplitudes $V_{l}$ and anomalous-hopping amplitudes $W_{l}$. The anomalous hopping terms are needed to generate the off-diagonal elements of the hybridization function.

The Anderson Hamiltonian can be implemented in the Fock basis, and the corresponding solution can be obtained by exact diagonalization of BDMFT~\cite{Appen_georges96}. After diagonalization, the local Green's function, which includes all the information about the bath, can be obtained from the eigenstates and eigenenergies
in the Lehmann-representation
\begin{eqnarray} \label{green_funtion}
G_{\rm imp}^1 (i \omega_n) &=& \frac{1}{Z} \sum_{mn} \langle m | \hat d | n\rangle \langle n | \hat d^\dagger | m \rangle \frac{e^{- \beta E_n} - e^{-\beta E_m}}{E_n - E_m + i \hbar } + \beta \Phi \Phi^\ast\\
G_{\rm imp}^2 (i \omega_n) &=& \frac{1}{Z} \sum_{mn} \langle m | \hat d | n\rangle \langle n | \hat d | m \rangle \frac{e^{- \beta E_n} - e^{-\beta E_m}}{E_n - E_m + i \hbar \omega_n} + \beta \Phi \Phi.
\end{eqnarray}

Integrating out the orbitals leads to the same effective action as in Eq.~(\ref{eff_action}), if the following identification is made
\begin{eqnarray}
\boldsymbol{\Delta} (i\omega_n)  & \equiv & t^2 {\sum_{\langle 0i\rangle ,\langle 0j \rangle }}^\prime\mathbf G^{(0)}_{ij}(i\omega_n),
\end{eqnarray}
where $\Delta^1(i \omega_n)  \equiv -\sum_l\Big(\frac{V_{l}V^\ast_{l}}{\epsilon_l-i\omega_n} + \frac{W^\ast_{l}W_{l}}{\epsilon_l+i\omega_n}\Big)$, $\Delta^2(i \omega_n) \equiv  -\sum_l\Big(\frac{V_{l}W^\ast_{l}}{\epsilon_l-i\omega_n} +
\frac{W^\ast_{l}V_{l}}{\epsilon_l+i\omega_n}\Big)$, and $\sum^\prime$ means summation only over the nearest neighbors of the "impurity site".

In next step, we make the approximation that the lattice self-energy $\Sigma_{i,\rm lat}$ coincides with the impurity self-energy $\Sigma_{i,\rm imp}$, which is obtained from the local Dyson equation
\begin{equation}
 \Sigma_{i,\rm imp} (i \omega_n) =  \left( \begin{array}{cc}
                          i \omega_n +\mu + \Delta^1 & \Delta^2 \\
                         \Delta^{2\ast} \ & -i \omega_n + \mu +\Delta^{1\ast}
                         \end{array}\right) - G^{-1}_{ii,\rm imp} (i \omega_n).
 \label{eq:localdysonequation}
\end{equation}
The real-space Dyson equation takes the following form:
\begin{equation}
 G^{-1}_{ij, \mathrm{latt}} (i \omega_n) =  \!\left( \begin{array}{cc}
                          \!\!\left(i \omega_n \!+\!\mu\!  -\! \Sigma_{i,\rm lat}^{11}\right)\delta_{ij}\!+ \!h_{ij}& \!\!-\Sigma_{i,\rm lat}^{12} \delta_{ij}\\
                         \!\!- \Sigma_{i,\rm lat}^{21} \delta_{ij}\ & \!\!\left(\!-\!i \omega_n \!+\! \mu \!-\! \Sigma_{i,\rm lat}^{22}\right)\delta_{ij}\!+\!h_{ij}\!\!
                         \end{array}\right).
                         \label{eq:realspacedyson}
\end{equation}
Here, the self-consistency loop is closed by Eq.~(\ref{green_funtion})-(\ref{eq:realspacedyson}), and this self-consistency loop is repeated until the desired accuracy for values of parameters $\epsilon_l$, $V_l$ and $W_l$ and superfluid order parameter $\Phi$ is obtained.

\section{Ground state phase diagram for $\phi = \pi/2$}
\begin{figure}[h]
	\includegraphics*[width=0.5\linewidth]{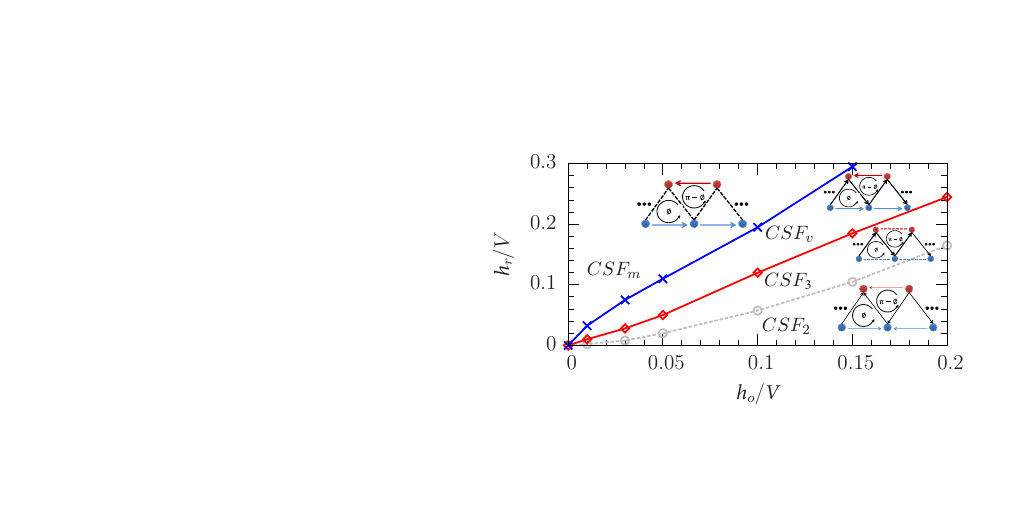}
	\caption{(Color online) Phase diagram of the quantum system with interacting spinwave states in a momentum-space lattice with the flux $\phi=\pi/2$ and filling factor $N_{\rm tot}/L_{\rm lat}=0.125$.	}
	\label{phase_0.5}
\end{figure}
Without two-body interactions, or with single Rydberg dressing, there are two special case for the flux $\phi=0$ and $\phi=\pi/2$. For the case $\phi=0$, the phenomena are trivial, and the system does not support edge currents in the absence (presence) of Rydberg long-range interactions. For the case $\phi=\pi/2$, the band structure of the lattice system is actually a double-valley well with two degenerate band minima connected by time-reversal symmetry, indicating that more ground states appear in the case. Here we choose the parameters: the filling factor $N_{\rm tot}/L_{\rm lat}=0.125$ ($L_{\rm lat}$ being the lattice size) and $\phi=\pi/2$. We observe there are four stable phases in the diagram with different types of ground state edge currents, including CSF$_2$ and CSF$_3$ with currents on the rung but with a suppressed global current on both ladders with $\bar{J}_{AA}=\sum_i J^i_{AA}/L_{\rm lat}\approx0$ and $\bar{J}_{BB}=\sum_i J^i_{BB}/L_{\rm lat}\approx 0$, CSF$_m$ with currents only on the ladders, and CSF$_v$ with currents on both ladders and rungs. The physical reason of the suppressed edge currents of the CSF$_2$ and CSF$_3$ phases is that, $J^i_{AA}\approx 2h_on_i{\rm sin}(\phi-2kx_A)\approx0$ and $J^i_{BB}\approx -2h_on_i{\rm sin}(\phi-2kx_B)\approx0$ with $\phi=\pi/2$, $n_i$ being the filling at site $i$, and condensing at $2kx_A\approx\pi/2$ and $2kx_B\approx-\pi/2$, respectively.

%\section{Density distribution in the ground state for $\phi=\pi/4$.}	
\section{Density distribution in real space and momentum space with single Rydberg dressing}
\begin{figure}[t]
	\includegraphics*[width=0.55\linewidth]{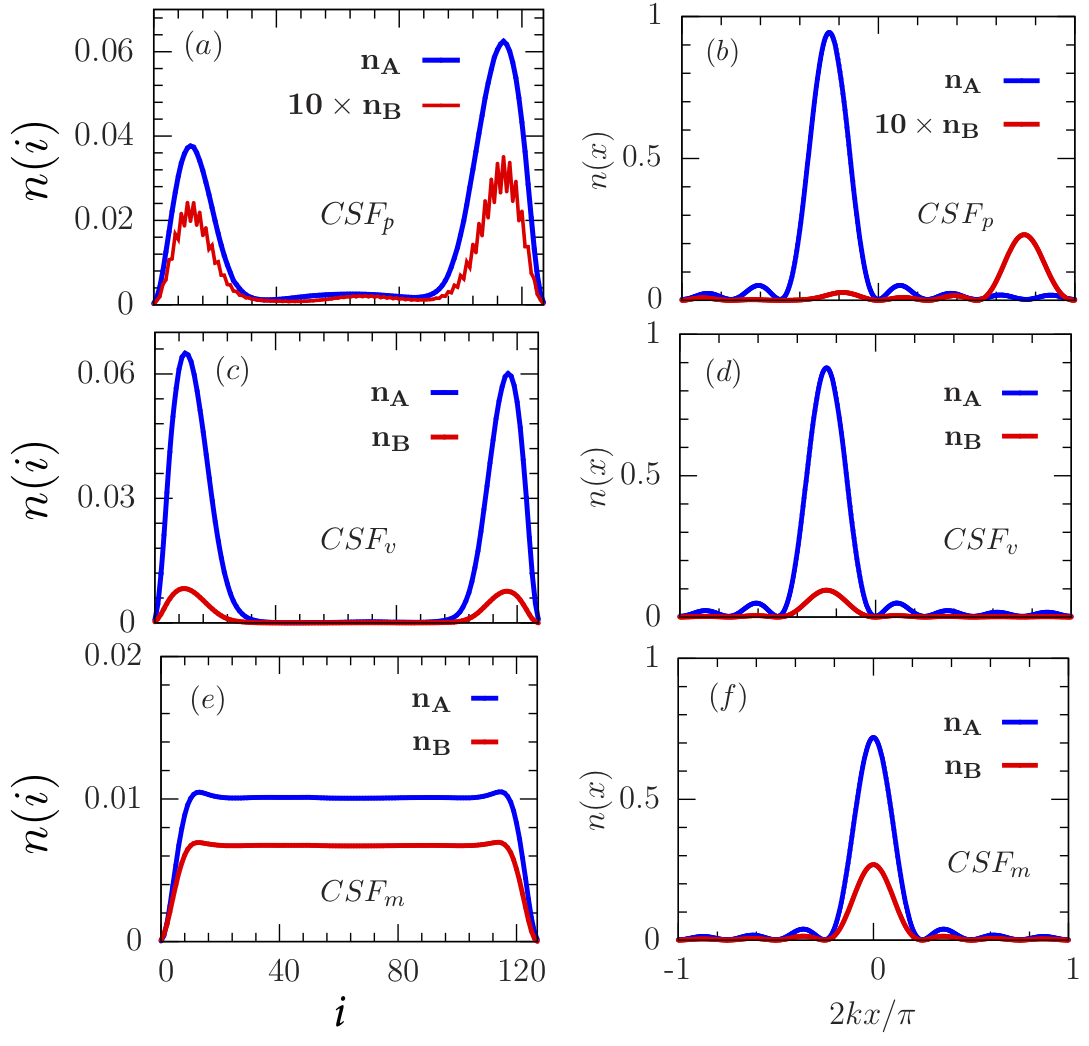}
	\vspace{-2mm}
	\caption{(Color online) {\textbf{Ground state density distribution of interacting spinwaves with open boundary condition}}. The momentum-space density distributions are shown for the CSF$_p$ (a), CSF$_v$ (c) and CSF$_m$ (e) phases.  The corresponding real-space density distributions are shown in (b), (d) and (f). Here the flux is $\phi=\pi/4$, the parameters $h_r/V$ and $h_o/V$ are $0.005$ and $0.2$ in the CSF$_p$ phase, $0.1$ and $0.2$ in the CSF$_v$ phase, and $0.3$ and $0.2$ in the CSF$_m$ phase, respectively.}
	\label{momentum}
\end{figure}

In the main text, we identified three phases in the case of single Rydberg dressing. The density distributions in momentum and real space are vastly different in the three phases. Here we will show both real-space and momentum-space density distributions in these phases, which might be observed directly through time-of-flight experiment. In the Meissner phase, the size of the vortex is infinite and the density is uniform in the momentum-space lattice. On the other hand, in the CSF$_p$ and CSF$_v$ phases, the system exhibits vortex structures, where the densities are distributed inhomogeneously and the vortexes are separated into different regions (the averaged inter-leg current $\sum_i J^i_{BB}/L_{\rm lat}\approx 0$ and $\neq 0$ for the CSF$_p$ and CSF$_v$ phases, respectively).

Actually, we have observed that there are three different kinds of band structures in the noninteracting system, as shown in Fig.~\ref{fig:system}(d), where band minima are localized at zero (CSF$_m$), finite-value (CSF$_v$), and doubly degenerate points (CSF$_p$) in real space~\cite{PhysRevLett.122.023601}. The interacting system also supports three types of condensation in real space, and the resulting phenomena are that, in the CSF$_p$ phase, the maximal density for the $|a\rangle$ and $|b\rangle$ states, localized at $x_A$ and $x_B$, respectively, separates by about $2kx\approx \pi$ [$J^i_{AA}\propto h_o{\rm sin}(\phi-2kx_A)>0$ and $J^i_{BB}\propto -h_o{\rm sin}(\phi-2kx_B)>0$ with $\phi=\pi/4$, as shown in Fig.~\ref{momentum}(b) for $x_A$ and $x_B$, and more discussions on the anti-chiral current are shown in the next section], and in the CSF$_v$ phase, peak positions of the two states are identical in real space condensing at nonzero value. In the CSF$_m$ phase, however, the maximal value of the density in the two states is centered at $2kx=0$ in real space. This indicates that we can directly identify the CSF$_p$ phase through the real-space distribution.

\section{CSF$_p$ phase without two-body interactions}
When the two-body interaction is vanishing, the Hamiltonian of an ensemble of atoms in real space is given by
\begin{eqnarray}
H=\sum_j2h_0\cos(2kx_j-\phi)(|b_j\rangle\langle b_j| - |a_j\rangle\langle a_j|) -2h_r(|a_j\rangle\langle b_j + |b_j\rangle\langle a_j|),
\end{eqnarray}
where $j$ is the index of atoms. The eigenvalues of the Hamiltonian are $E_{\pm}=\pm\sqrt{2} \times\sqrt{h_0^2[1 + \cos (4 kx_j-2 \phi )]+h_r^2 [1 + \cos (2
	kx_j)]}$.

In the limit $h_r\to 0$ the two eigenvalues as a function of $k x_j$ will cross at $kx_j= \phi/2 + \pi/4$. We can also find that when $k_1x_j=\phi/2$ (for state $|a\rangle$) and $k_2x_j=\phi/2 + \pi/2$ (for state $|b\rangle$), the eigenenergies are local minimal. The current $\bar{J}_{AA}=\frac{2h_0}{L_{\rm lat}}\sum_j\sin(\phi - 2kx_j)n_j^{(a)}$ and $\bar{J}_{BB}=-\frac{2h_0}{L_{\rm lat}}\sum_j\sin(\phi - 2kx_j)n_j^{(b)}$ are zero in this limit.

Now focusing on the case $\phi=\pi/4$ and turning on the coupling $h_r$ between the two states, the two local minimal points are coupled. When $h_r\ll h_0$, the two local minimal points are still nearly degenerate. The minimal points are shifted slightly with respect to $k_1$ and $k_2$.  We can expand the lower branch $E_-$ of the eigenenergy around $h_r\sim 0$ up to second order and find the shifts, $\Delta k_1x_j\approx -h_r^2\sin\phi/4h_0^2$ and $\Delta k_2 x_j\approx -h_r^2\sin\phi/4h_0^2$. The current $\bar{J}_{AA}\approx \frac{2h_0}{L_{\rm lat}}\sum_j\sin(-\Delta k_1)n_j^{(a)} \approx \frac{2h_0}{L_{\rm lat}}\sum_j |\Delta k_1|n_j^{(a)} >0$. Similarly, we find that $\bar{J}_{BB}>0$. Here $\bar{J}_{AA}>\bar{J}_{BB}$ because the state $|b\rangle$ is weakly occupied in the ground state, due to small $h_r$. This explains the results shown in the main text.

\section{Sliding phases in the strongly interacting regime}
\begin{figure}[h]
	\includegraphics*[width=0.5\linewidth]{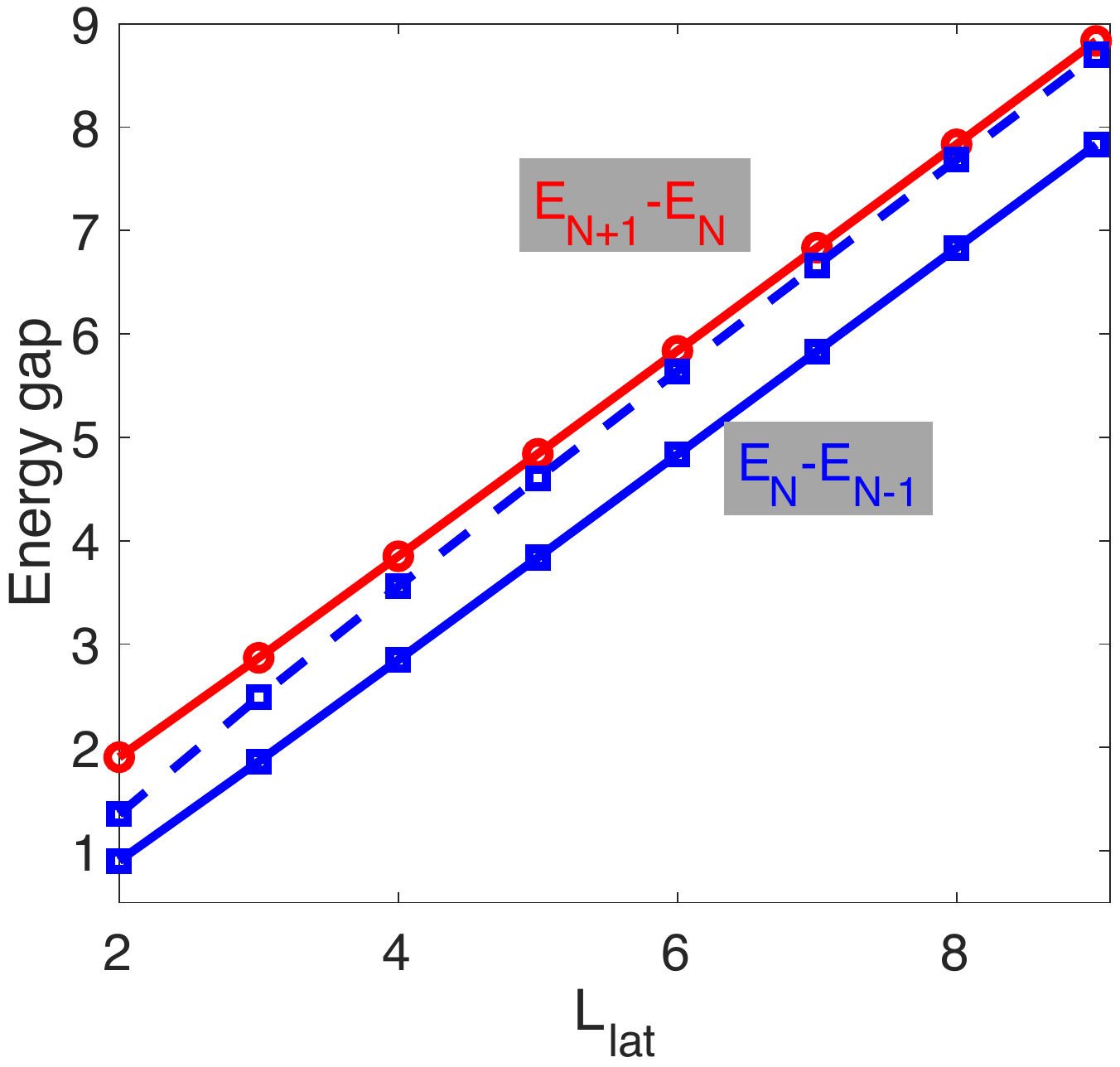}
	\vspace{-2mm}
	\caption{(Color online) Ground state energy gap by adding one particle $\Delta E^+_N = E_{N+1}-E_N$ (circle) and removing one particle $\Delta E^-_N=E_N-E_{N-1}$ (square). Without hopping, $\Delta E^+_N= N\Delta E^-_N/(N-1)$. In the presence of hopping, we plot $N\Delta E^-_N/(N-1)$ (dashed), which approaches to $\Delta E^+_N$ when the number of total particles $N$ increases, with $E_N$ denoting the ground state energy of the system. Here we consider the ground state energy of the A-leg.  The flux is $\phi=0.25\pi$, the other parameters are $h \equiv h_r/V=0.1$.}
	\label{app_egap}
\end{figure}
%\bblue{For a bosonic system with nearest-hopping $h$ and onsite interaction $U$ in optical lattices, which is described by the lowest-band Bose-Hubbard model $H=-h\sum_{\langle i, j \rangle} b^\dagger b  + V/2\sum_i n_i(n_i-1)$. The relative bandwidth of the system is $\Delta_D \approx h/Un_i(n_i-1)$, where it remains a finite value in the thermodynamical limit. Here, $b^\dagger$ denotes the creation operator of bosons at site $i$, and $n_{i\sigma}\equiv b^\dagger_{i}b_i$ the atomic density at site $i$.}

The unique properties of the sliding phase can be understood by considering a single chain. In the following we will consider the A-leg with $N$ excitation from the atomic hyperfine ground state as an example. In the momentum space, all the excitations are coupled by the infinite long-range Rydberg interactions. The interaction energy $E_{\rm int}= V N(N -1)/2$ regardless of the particle distribution. The energy gap by adding one particle $\Delta E^+_N =E_{N+1}-E_N=NV$ and removing one particle $\Delta E^-_N =E_N-E_{N-1}=(N-1)V$ depends on particle numbers linearly, where $E_N$ denotes the ground state energy of the system. This is approximately valid even in the presence of a weak hopping. To verify this, we have carried out exact diagonalization (ED) calculations for unit filling $N=L_{a}$. To calculate the gap, we have considered system size up to $N=10$ and $L_{a}=9$, whose Hilbert space is 43758. Our numerical results show that the energy gap is indeed proportional to particle numbers (see Fig.~\ref{app_egap}). On the other hand, the ratio of the two gaps $\Delta E^+_N/\Delta E^-_N= N/(N-1)$ when the hopping vanishes. The numerical simulation shows that the ratio is approximately correct (Fig.~\ref{app_egap}). The deviation in the numerical calculation is due to the finite size effect.

Moreover, we can understand the existence of the sliding phase  from the relative bandwidth of the system, i.e. the ratio between the hopping and interaction energy. First, the maximal value of the hopping energy is $h[\sqrt{N(N+1)}+ \sqrt{(N-1)(N+2)}]/2$, corresponding to a configuration where $(N-1)/2$ and $(N+1)/2$ atoms locate at two neighboring sites. In the thermodynamic limit, the relative bandwidth is estimated as
\begin{equation}
\Delta_D \approx \frac{ hN}{E_{\rm int}} \propto \mathcal{O}(1/N).
\end{equation}
This power decay indicates that the relative bandwidth vanishes for the Rydberg-dressed system in the thermodynamic limit. As a result, a sliding phase emerges in the strongly interacting limit. As a comparison, we note that the relative bandwidth in the Mott insulator state of the Bose-Hubbard model, described by $H=-h\sum_{\langle i, j \rangle} b^\dagger b  + V/2\sum_i n_i(n_i-1)$, is $\Delta_D \propto 1/n(n-1)$ with filling $n$ (i.e. the mean number of atoms in a site).  The corresponding bandwidth is constant in the thermodynamic limit.

\begin{figure}[t]
	\includegraphics*[width=1\linewidth]{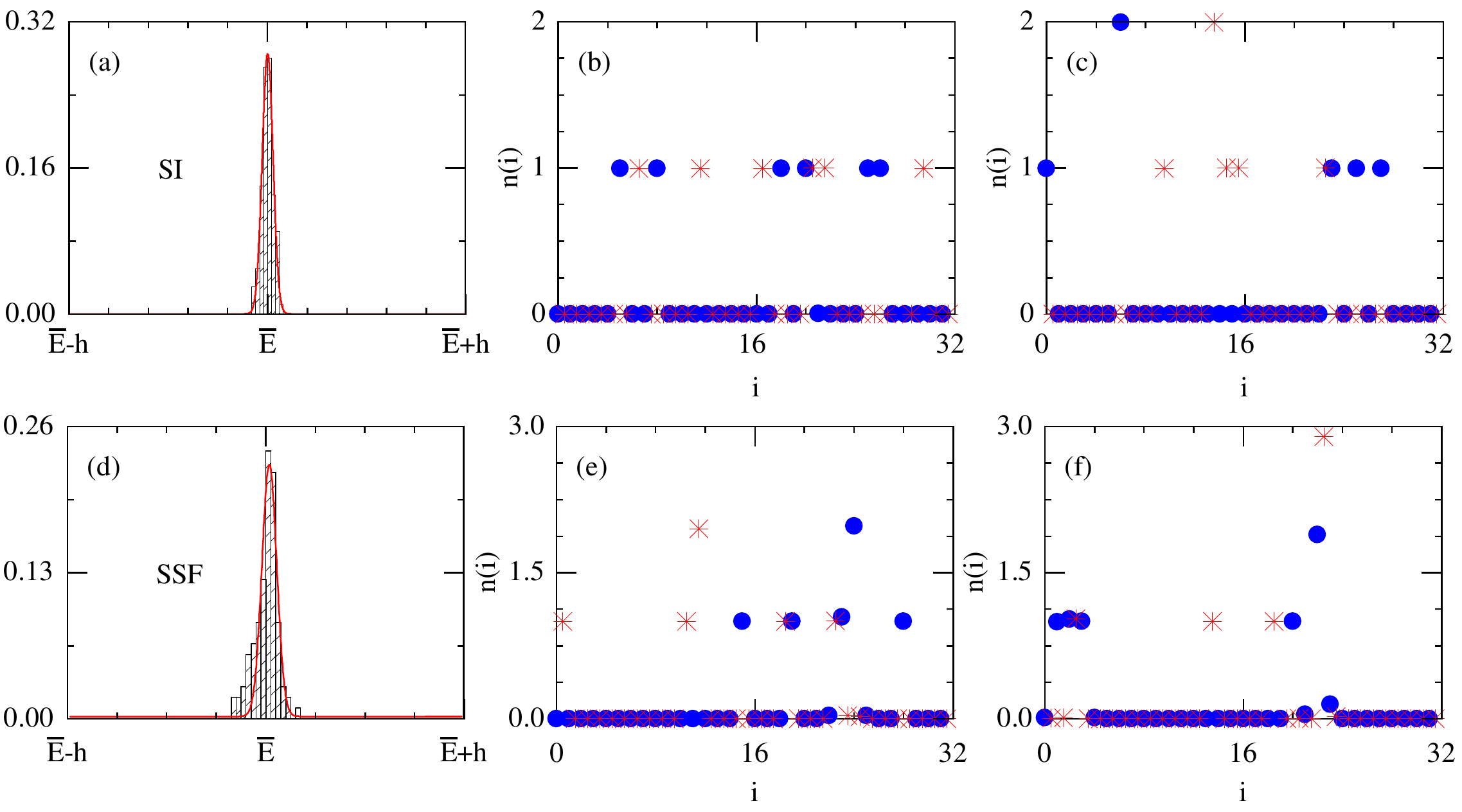}
	\vspace{-2mm}
	\caption{(Color online) {\textbf{Probability density histograms of the system's energies for the sliding phases in the strongly interacting regime with both chains dressed to Rydberg states}}. (a)(d) Distribution of average energy per particle $\bar E$ for 100 times simulations, based on dynamical mean-field theory, where the red line is a Gaussian fitting. Here, we observe the calculated energies are localized around the center (compared to hopping amplitudes) for different density distributions in momentum-space lattice, indicating sliding nature of the many-body states. As shown in (b)(c) and (e)(f), two examples of density distributions are shown for two different numerical simulations, where the blue circle and red cross denote particles in A-leg and B-leg, respectively. Here, the flux is $\phi=0$, the other parameters $h \equiv h_r/V=h_o/V=0.05$ (a-c), and $h \equiv h_r/V=h_o/V=0.2$ (d-f).}
	\label{app_sliding}
\end{figure}
It becomes difficult to carry out ED calculation for larger systems, such as two coupled chains with larger $L_{\rm lat}$. Here we calculate energy distributions for different numerical simulations via the dynamical mean-field theory. The ground state energies, with different density distributions, localize in a small region (compared to hopping amplitudes), indicating the sliding nature of the quantum many-body states at finite hopping amplitudes, as shown Fig.~\ref{app_sliding}. The broadening of the spectra is largely caused by the kinetic energy of different configurations of the finite-size system [see Fig.~\ref{app_sliding}(b), (c), (e) and (f)]. As examples, we also show density distributions in momentum space from two different numerical simulations [Fig.~\ref{app_sliding}(b) and (c) in SI, and Fig.~\ref{app_sliding}(e) and (f) in SSF region] but with the same parameters. In different simulations, the particles distribute randomly in each ladder, as shown in Fig.~\ref{app_sliding}(b)(c) and (e)(f). Experimental observation of such distribution could be a direct evidence of the sliding phase.

\section{Quantum Zero dynamics with effective decay}
\begin{figure}[h]
	\includegraphics*[width=0.59\linewidth]{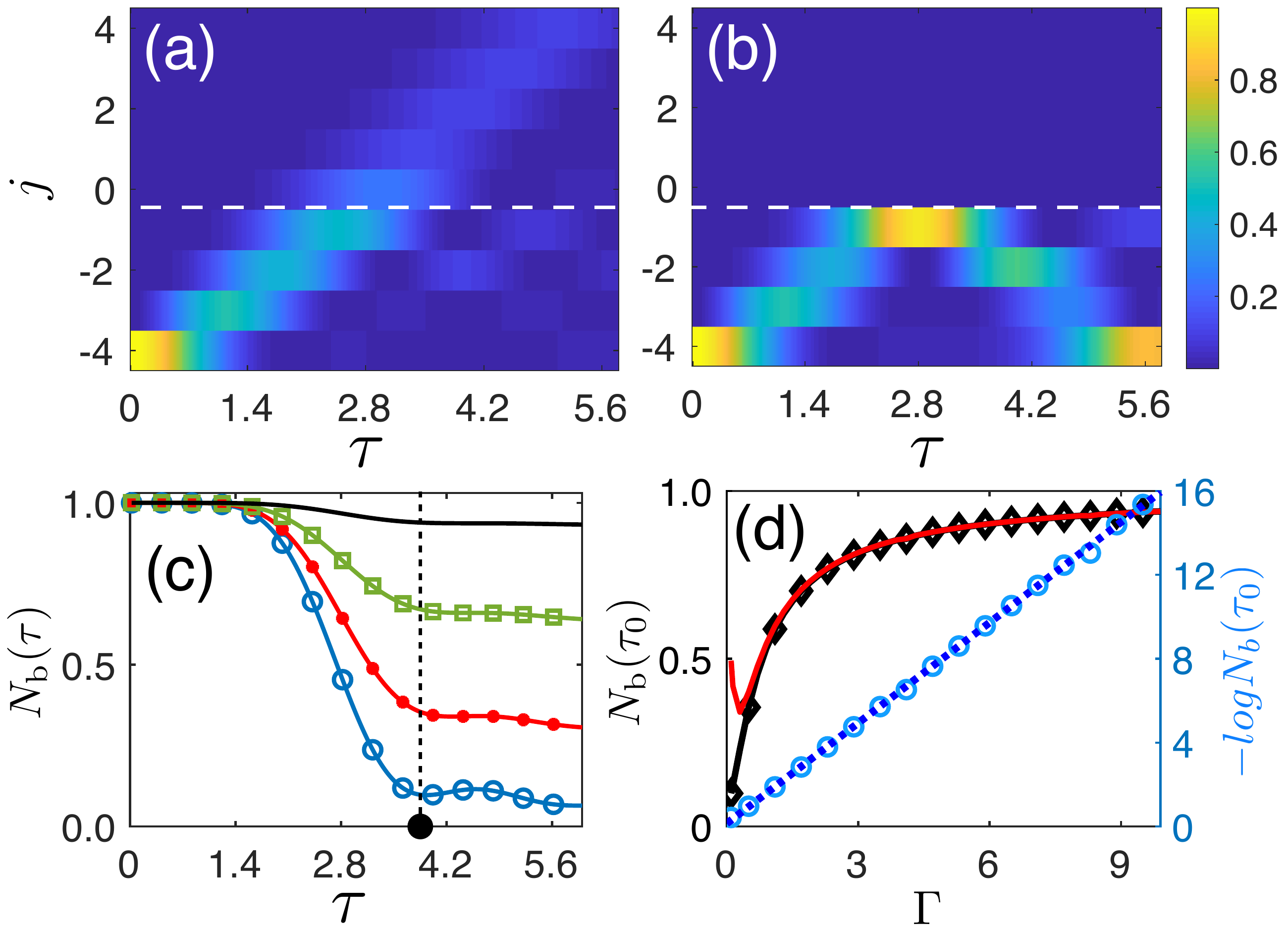}
	\caption{(Color online) {\textbf{Zeno dynamics in the dissipative regime}.} Loss (a) and reflection (b) of the spin wave. Without decay of the spin wave, population propagates along the lattice and losses at the zeroth and first sites. Strong dissipation in site $|n^{(a)}_0\rangle$ and $|n^{(b)}_{1}\rangle$ reflects the population.  In (a) $\Gamma_0=\Gamma_{1}=\Gamma=0.01$ and (b)  $\Gamma_0=\Gamma_{1}=\Gamma=2.0$, while decay in other sites being negligible.  (c) The remaining population $N_b(\tau)$ (from the initial site to the middle of the lattice, denoted by the dashed line.) is show at different times for decay rate $\Gamma_0=\Gamma_{1}=\Gamma$ from 0.1 (circle), 0.5 (star), and 1.5 (square) to 10 (solid). (d) Saturation values (diamond) of the remaining population at $\tau=h_o t = 3.8$. When the decay is strong, the total population (red solid) is identical the reflected population, i.e. fully reflected. The remaining population depends on $\Gamma$ exponentially (circles). The dashed line is the fitting of $-\log N_b(\tau_0)$. Other parameters are $\phi = 0$, $h_0=h_r=0.2$.}
	\label{fig:zeno}
\end{figure}
When the long-lived hyperfine states are replaced with low-lying decay states, we realize a superradiance lattice~\cite{wang_superradiance_2015}, or incorporate an effective decay in the present setup. To highlight the collective decay, we study dissipative dynamics of the A-leg solely and turn off the dressing laser.  It could be realized by coupling the level $|a\rangle$ with a strong classical field off-resonantly. Spontaneous Raman process happens in a rare probability, which depends on the detuning. It's a reversal process of the state initialization as we mentioned above.

Without two-body interactions, we will study a single excitation in the A-leg, whose dynamics is now described by a master equation \begin{equation}
\dot{\rho}=-i[H,\rho]+\sum_{j}\Gamma_j(a_j\rho a^{\dagger}_j-\{a^{\dagger}_ja_j,\rho\}),
\end{equation}
where $ H_A = -\sum_i(h_0e^{i\phi} a_i^{\dagger}a_{i+1} + \rm{H.c})$. $\Gamma_j$ are decay rate of the $j$-th site. Propagation of the spinwave depends on hopping $h_0$ and decay rate $\Gamma_j$. The latter is site (momentum) dependent for a superradiance lattice. With this consideration, $\Gamma_j\neq 0$ only for phase matched spinwave components~\cite{Svidzinsky_08}, i.e. at sites with index $j=0$ and $1$ [see Fig.~\ref{fig:system}(b)].

We consider a single excitation initially occupies the state $|n^{(a)}_{-4}=1\rangle$ and propagates to the middle of the lattice. One typically expects that the propagation is coherent away from the central sites, and becomes dissipative once approaching to the decaying sites. This is true when the decay rate $\Gamma_j$ is small, where a large fraction of the population will be lost [Fig.~\ref{fig:zeno}(a)]. The dynamics changes qualitatively when $\Gamma_j$ is large, where the population is largely reflected at the zeroth site  [Fig.~\ref{fig:zeno}(b)(c)].

We attribute the reflection to the quantum Zeno effect~\cite{scully_quantum_1997}. Stronger decay behaves similar to a frequent measurement of occupations in the zeroth site. It prohibits the occupation of its neighboring sites from hopping to the initially not occupied center sites. As a result, the loss occurs at an effective, smaller rate $h_o/\Gamma_0^2$~\cite{daley_quantum_2014}. To illustrate this, we consider the simplest model which contains only the two sites with index $j=-1$ and $j=0$. The initial state is $|\psi(0)\rangle=|n_{-1},n_0\rangle$ with $n_{-1}=1$ and $n_0=0$ (the subscript indicates the site index.). This state couples to the state $|\phi\rangle=|0,1\rangle$, which will decay at a rate $\Gamma_0^{(0)}$. The effective decay rate can be estimated through analyzing the non-Hermitian Hamiltonian~\cite{daley_quantum_2014}, $H_e=H_c+ H_d=(h_0e^{i\phi}a_{-1}a_0^{\dagger} +\text{H.c})-i{\Gamma}_0/2a^{\dagger}_0a_0$, where the coupling $H_c=(h_0e^{i\phi}a_{-1}a_0^{\dagger} +\rm{H.c})$ and diagonal Hamiltonian $H_d=-i\Gamma_0/2a_0^{\dagger}a_0$. The energy of the initial state in the presence of the coupling can be solved through the second order perturbation~\cite{daley_quantum_2014}
\begin{equation}
E_0^{(2)} = \frac{\langle \psi(0)|H_c|\phi\rangle\langle \phi| H_c|\psi_(0)\rangle}{-H_d}=\frac{h_0^2}{i\Gamma_0/2} = -\frac{2ih_0^2}{\Gamma}.
\end{equation}
The initial state evolves according to $|\psi(t)\rangle = \exp(-iE_0^{(2)}t)|\psi(0)\rangle$. Hence the remaining probability of the initial state is
\begin{eqnarray}
P=\langle \psi(t)|\psi(t)\rangle = \exp\left(-\frac{4h_0^2t}{\Gamma_0}\right).
\end{eqnarray}
The initial state decays at an effective rate $\Gamma_{\rm{eff}}= 4h_0^2/\Gamma_0$, which decreases with increasing $\Gamma_0$.

We numerically calculate the remaining population as a function of $\Gamma_0$ at time $\tau=h_0 t = 3.8$, when the reflection occurs. As shown in Fig.~\ref{fig:zeno}(d), the effective decay rate is linearly proportional to $\sim  h_o/\Gamma_0^2$, confirming the analytical prediction.

%\section{Effective decay rate in the transport of spinwave states}
%We study the transportation of spinwave states by considering only the A-leg. The dynamics is governed by a master equation
%\begin{equation}
%\dot{\rho}_A=-i[H_A,\rho_A]+\sum_j \Gamma_ja_j\rho_Aa_j^{\dagger} + 1/2\{a_j^{\dagger}a_j,\rho_A\},
%\end{equation}
%where the Hamiltonian reads $ H_A = -\sum_i(h_0e^{i\phi} a_i^{\dagger}a_{i+1} + \rm{H.c})$.
%
%Decay rates of the central sites ( index $j=0,1$) are large and can be neglected at other sites, i.e. $\Gamma_j>0$ for $j=0,1$ and $\Gamma_j=0$ otherwise. Far away from the center, the excitation propagates along the leg according to Hamiltonian $H_A$. Once approaching the middle sites, significant excitation loss takes place, when it hops from the nondecay neighboring site to the middle two sites. For situations shown in the main text, the hopping is from site $j=1$ to $j=0$.

\section{Experimental proposal for realizing strongly correlated phenomena}
Chiral edge states can be realized in momentum space, either using mechanical momentum states of cold atoms~\cite{gadway_atom-optics_2015,an2017direct,an_correlated_2018,an_engineering_2018,Meier929}, or collective excitation formed by collective excitations of electronic states~\cite{scully_directed_2006}. A unique advantage of the latter is that thermal resistant edge states can be probed, since the momentum-space lattice~\cite{wang_superradiance_2015,PhysRevLett.120.193601} of collective atomic excitations is immune to the motional entropy of atoms. The first proof-of-principle experiment has demonstrated chiral edge currents at the \textit{room temperature}~recently~\cite{PhysRevLett.122.023601}. However, electronic excited states suffer from fast spontaneous decay, inducing a steady state in the pump-dissipative system and destroying the $quantum$ nature of the system. A clean {\it quantum} system in the momentum-space lattice in the presence of strong interactions is required to simulate strongly correlated phenomena.
\subsection{Realization for the Hamiltonian (1) without dissipation}\label{sec:proposal}
We can get rid of the radiative dissipation by selecting three hyperfine spin states in ground levels of $^{87}{\rm Rb}$ with $|g\rangle$ being $|5^2S_{1/2},F=1,m=1\rangle$, $|a\rangle$ being $5^2S_{1/2},F=1,m=-1\rangle$, and $|b\rangle$ being $|5^2S_{1/2},F=2,m=-2\rangle$. The atomic level scheme and the configuration of the coupling beams are plotted in Fig.~\ref{fig:scheme}(a)(b). The spin states are split by a bias magnetic field.  The inter-leg coupling $h_r$ in purple is realized by Raman interaction which composed by the two fields around D1 line (in purple), while the one in blue is realized by a 6.8GHz microwave field. Since microwave field transfers negligible momentum, the MSL is slightly tilted in momentum space, as shown in Fig.~\ref{fig:scheme}(c). The intra-leg coupling $h_0$ is implemented by the standing waves around D2 line, whose frequency is set in the middle point between $a$ and $b$ levels to introduce the $\pi$ phase shift between the intra-leg couplings of A- and B-legs. The phase $\phi$ in Ham. (1) can be  controlled by manipulating the phase of the microwave and optical driving fields. Note here that, the small difference between the wave vectors of the standing wave is negligible. In the low-excitation regime, these atomic spinwave states are described by bosons~\cite{Fleischhauer_00}.

Here we choose Rydberg 60$S$ state as an example, whose lifetime is 102 $\mu$s. With the detuning $\Delta_d = 8$ MHz and Rabi frequency $\Omega_R=3$ MHz of the dressing laser, the effective lifetime in the Rydberg dressed state is 2.9 ms. We obtain $r_c=4.54\, \mu$m and $V=158.2$ kHz. This large soft-core radius $r_c$ leads to a short-range interaction in momentum space, as it is far larger than the wavelength of the standing wave laser ($\sim 780$ nm). The inter- and intra-leg coupling strengths $h_r$ and $h_o$ can vary in a large parameter regimes. For example, we choose the parameters $\Delta_{zeeman}=50\,{\rm MHz}$ being the Zeeman splitting between $|g\rangle$ and $|a\rangle$, $\Delta_{r}=\Delta_a=\Delta_b=3.4\,{\rm GHz}$ being the detunings [see details in Fig.~\ref{fig:scheme}(a)], $\Omega_s=40\,{\rm MHz}$  being the Rabi frequencies of optical fields with $s=0,1,2$, and $\Omega_{mw}=500\,{\rm kHz}$ being the effective Rabi frequency of the microwave field. The corresponding $h_r$ and $h_0$ are $\sim 50\, {\rm kHz}$ ($\gg {\rm E_r} \approx 4\, \rm {kHz}$, with ${\rm E_r}$ being the recoil energy), which are larger enough to observe coherent dynamics in microsecond timescale.

\begin{figure}[h!]
	\includegraphics[width=0.6\linewidth]{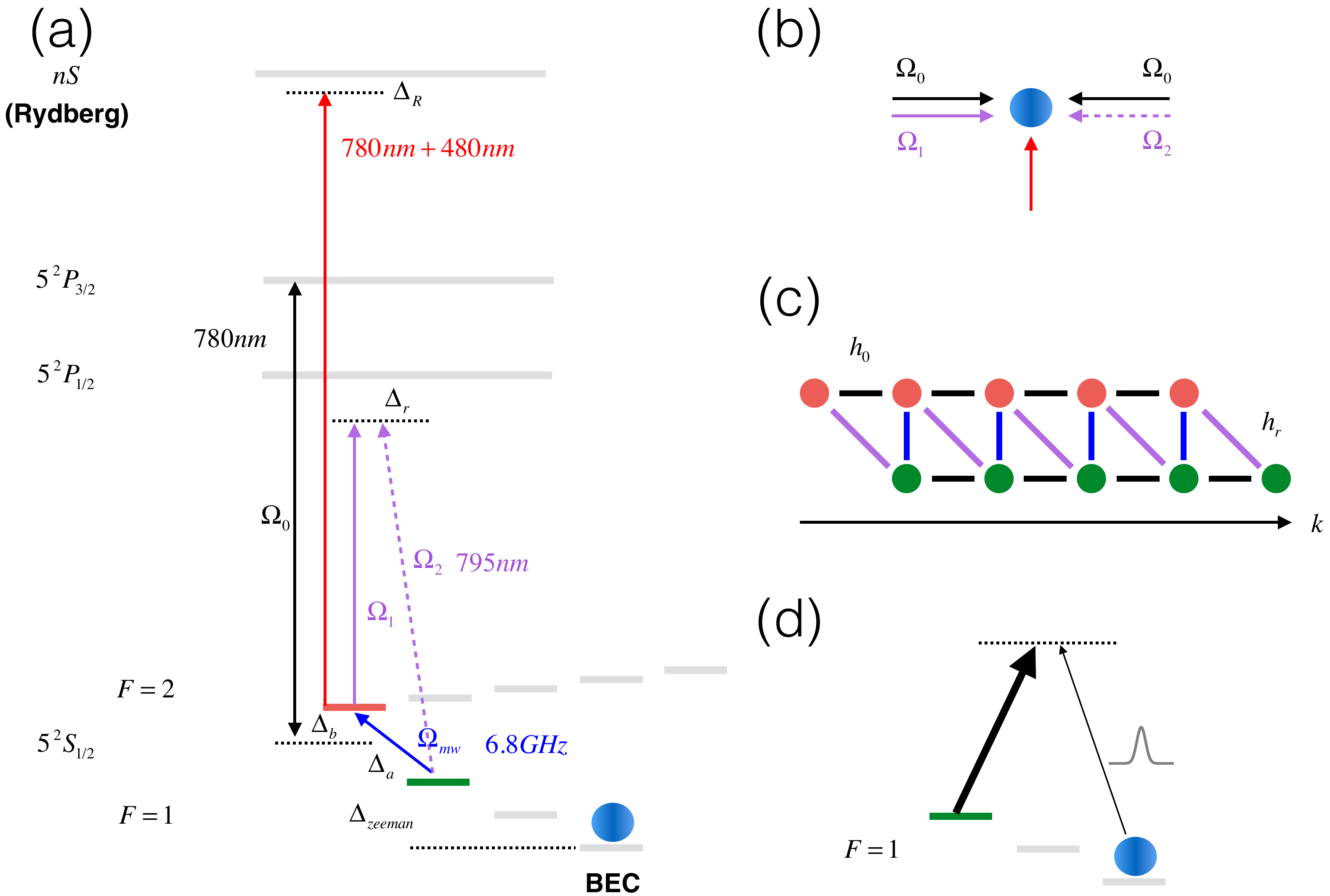}
	\caption{(Color online) \textbf{Experimental setup}. (a) The atomic level scheme. (b) The configuration of the coupling optical fields. (c) The momenmum space lattice. (d) The Raman coupling to prepare the initial state.}\label{fig:scheme}
\end{figure}

\subsection{Loading the  excitations and measurement}

To observe the dynamics in MSL, we need to initialize excitations in the A-leg with a post-selection process. The BEC is prepared in ground state before turning on the Ham.~(\ref{eff_Ham}). Then we apply a strong classical field (thick) and a single photon (thin) in Fig.~\ref{fig:scheme}(d), forming a Raman coupling. When the single photon is not observed by a detector on its incident direction, we know one excitation is loaded in level $a$ \cite{scully_directed_2006}. Since the coherent time of level $a$ is long enough, we can repeat the  process twice to prepare the two-excitation state in zeroth site in A-leg. For the Zeno dynamics, we need to prepare a single excitation on the $n$th site  in A-leg and introduce an effective decay to the zeroth one (see more details in the next subsection). After the single excitation is loaded into the zeroth site, we can apply  two $\pi$-pulses of blue and purple $h_r$ couplings in a sequence. Such a pulse pair transport the excitation from  the zeroth site to the first site in the A-leg. We can repeat the process for $n$ times to finish the initialization.\cite{wang2014heisenberg}.

We can also pump the excitation to the MSL when the Hamiltonian is on.  By tuning the frequency of the pumping  microwave field to the energy of the ground state in MSL, we can excite a specific state with high fidelity since the state width is very narrow. To prepare the total excitations $N_{\rm tot}\ll N_{\rm BEC}$, the pulse area of the pumping microwave field is roughly estimated as $\sqrt{N_{\rm BEC}}\Omega_pt=\pi N_{\rm tot}$, where $\Omega_p$ is the effective Rabi frequency of the pumping microwave field, $\sqrt{N_{\rm BEC}}$ is the collective enhancement of $N_{\rm BEC}$ atoms, and $t$ is the pulse duration.

By measuring the probability distribution of the $a$- and $b$-level atoms in momentum space via time of flight imaging, we can obtain the distribution of the excitation in MSL, which is expected to show the strongly correlated phenomena, {\it e.g.} ground state chiral current \cite{Atala2014Observation}, excitation blockade, and Zeno dynamics.

\end{widetext}

\bibliography{references}

\begin{thebibliography}{90}%
\makeatletter
\providecommand \@ifxundefined [1]{%
 \@ifx{#1\undefined}
}%
\providecommand \@ifnum [1]{%
 \ifnum #1\expandafter \@firstoftwo
 \else \expandafter \@secondoftwo
 \fi
}%
\providecommand \@ifx [1]{%
 \ifx #1\expandafter \@firstoftwo
 \else \expandafter \@secondoftwo
 \fi
}%
\providecommand \natexlab [1]{#1}%
\providecommand \enquote  [1]{``#1''}%
\providecommand \bibnamefont  [1]{#1}%
\providecommand \bibfnamefont [1]{#1}%
\providecommand \citenamefont [1]{#1}%
\providecommand \href@noop [0]{\@secondoftwo}%
\providecommand \href [0]{\begingroup \@sanitize@url \@href}%
\providecommand \@href[1]{\@@startlink{#1}\@@href}%
\providecommand \@@href[1]{\endgroup#1\@@endlink}%
\providecommand \@sanitize@url [0]{\catcode `\\12\catcode `\$12\catcode
  `\&12\catcode `\#12\catcode `\^12\catcode `\_12\catcode `\%12\relax}%
\providecommand \@@startlink[1]{}%
\providecommand \@@endlink[0]{}%
\providecommand \url  [0]{\begingroup\@sanitize@url \@url }%
\providecommand \@url [1]{\endgroup\@href {#1}{\urlprefix }}%
\providecommand \urlprefix  [0]{URL }%
\providecommand \Eprint [0]{\href }%
\providecommand \doibase [0]{http://dx.doi.org/}%
\providecommand \selectlanguage [0]{\@gobble}%
\providecommand \bibinfo  [0]{\@secondoftwo}%
\providecommand \bibfield  [0]{\@secondoftwo}%
\providecommand \translation [1]{[#1]}%
\providecommand \BibitemOpen [0]{}%
\providecommand \bibitemStop [0]{}%
\providecommand \bibitemNoStop [0]{.\EOS\space}%
\providecommand \EOS [0]{\spacefactor3000\relax}%
\providecommand \BibitemShut  [1]{\csname bibitem#1\endcsname}%
\let\auto@bib@innerbib\@empty
%</preamble>
\bibitem [{\citenamefont {Klitzing}\ \emph {et~al.}(1980)\citenamefont
  {Klitzing}, \citenamefont {Dorda},\ and\ \citenamefont
  {Pepper}}]{Klitzing1980New}%
  \BibitemOpen
  \bibfield  {author} {\bibinfo {author} {\bibfnamefont {K.~V.}\ \bibnamefont
  {Klitzing}}, \bibinfo {author} {\bibfnamefont {G.}~\bibnamefont {Dorda}}, \
  and\ \bibinfo {author} {\bibfnamefont {M.}~\bibnamefont {Pepper}},\
  }\href@noop {} {\bibfield  {journal} {\bibinfo  {journal} {Physical Review
  Letters}\ }\textbf {\bibinfo {volume} {45}},\ \bibinfo {pages} {494}
  (\bibinfo {year} {1980})}\BibitemShut {NoStop}%
\bibitem [{\citenamefont {Thouless}\ \emph {et~al.}(1982)\citenamefont
  {Thouless}, \citenamefont {Kohmoto}, \citenamefont {Nightingale},\ and\
  \citenamefont {den Nijs}}]{PhysRevLett.49.405}%
  \BibitemOpen
  \bibfield  {author} {\bibinfo {author} {\bibfnamefont {D.~J.}\ \bibnamefont
  {Thouless}}, \bibinfo {author} {\bibfnamefont {M.}~\bibnamefont {Kohmoto}},
  \bibinfo {author} {\bibfnamefont {M.~P.}\ \bibnamefont {Nightingale}}, \ and\
  \bibinfo {author} {\bibfnamefont {M.}~\bibnamefont {den Nijs}},\ }\href
  {\doibase 10.1103/PhysRevLett.49.405} {\bibfield  {journal} {\bibinfo
  {journal} {Phys. Rev. Lett.}\ }\textbf {\bibinfo {volume} {49}},\ \bibinfo
  {pages} {405} (\bibinfo {year} {1982})}\BibitemShut {NoStop}%
\bibitem [{\citenamefont {Hasan}\ and\ \citenamefont
  {Kane}(2010)}]{hasan_colloquium:_2010}%
  \BibitemOpen
  \bibfield  {author} {\bibinfo {author} {\bibfnamefont {M.~Z.}\ \bibnamefont
  {Hasan}}\ and\ \bibinfo {author} {\bibfnamefont {C.~L.}\ \bibnamefont
  {Kane}},\ }\href {\doibase 10.1103/RevModPhys.82.3045} {\bibfield  {journal}
  {\bibinfo  {journal} {Reviews of Modern Physics}\ }\textbf {\bibinfo {volume}
  {82}},\ \bibinfo {pages} {3045} (\bibinfo {year} {2010})}\BibitemShut
  {NoStop}%
\bibitem [{\citenamefont {Senthil}(2015)}]{senthil_symmetry-protected_2015}%
  \BibitemOpen
  \bibfield  {author} {\bibinfo {author} {\bibfnamefont {T.}~\bibnamefont
  {Senthil}},\ }\href {\doibase 10.1146/annurev-conmatphys-031214-014740}
  {\bibfield  {journal} {\bibinfo  {journal} {Annual Review of Condensed Matter
  Physics}\ }\textbf {\bibinfo {volume} {6}},\ \bibinfo {pages} {299} (\bibinfo
  {year} {2015})}\BibitemShut {NoStop}%
\bibitem [{\citenamefont {Sinova}\ \emph {et~al.}(2015)\citenamefont {Sinova},
  \citenamefont {Valenzuela}, \citenamefont {Wunderlich}, \citenamefont
  {Back},\ and\ \citenamefont {Jungwirth}}]{sinova_spin_2015}%
  \BibitemOpen
  \bibfield  {author} {\bibinfo {author} {\bibfnamefont {J.}~\bibnamefont
  {Sinova}}, \bibinfo {author} {\bibfnamefont {S.~O.}\ \bibnamefont
  {Valenzuela}}, \bibinfo {author} {\bibfnamefont {J.}~\bibnamefont
  {Wunderlich}}, \bibinfo {author} {\bibfnamefont {C.~H.}\ \bibnamefont
  {Back}}, \ and\ \bibinfo {author} {\bibfnamefont {T.}~\bibnamefont
  {Jungwirth}},\ }\href {\doibase 10.1103/RevModPhys.87.1213} {\bibfield
  {journal} {\bibinfo  {journal} {Reviews of Modern Physics}\ }\textbf
  {\bibinfo {volume} {87}},\ \bibinfo {pages} {1213} (\bibinfo {year}
  {2015})}\BibitemShut {NoStop}%
\bibitem [{\citenamefont {Hansson}\ \emph {et~al.}(2017)\citenamefont
  {Hansson}, \citenamefont {Hermanns}, \citenamefont {Simon},\ and\
  \citenamefont {Viefers}}]{hansson_quantum_2017}%
  \BibitemOpen
  \bibfield  {author} {\bibinfo {author} {\bibfnamefont {T.~H.}\ \bibnamefont
  {Hansson}}, \bibinfo {author} {\bibfnamefont {M.}~\bibnamefont {Hermanns}},
  \bibinfo {author} {\bibfnamefont {S.~H.}\ \bibnamefont {Simon}}, \ and\
  \bibinfo {author} {\bibfnamefont {S.~F.}\ \bibnamefont {Viefers}},\ }\href
  {\doibase 10.1103/RevModPhys.89.025005} {\bibfield  {journal} {\bibinfo
  {journal} {Reviews of Modern Physics}\ }\textbf {\bibinfo {volume} {89}},\
  \bibinfo {pages} {025005} (\bibinfo {year} {2017})}\BibitemShut {NoStop}%
\bibitem [{\citenamefont {Bloch}\ \emph {et~al.}(2008)\citenamefont {Bloch},
  \citenamefont {Dalibard},\ and\ \citenamefont {Zwerger}}]{RevModPhys.80.885}%
  \BibitemOpen
  \bibfield  {author} {\bibinfo {author} {\bibfnamefont {I.}~\bibnamefont
  {Bloch}}, \bibinfo {author} {\bibfnamefont {J.}~\bibnamefont {Dalibard}}, \
  and\ \bibinfo {author} {\bibfnamefont {W.}~\bibnamefont {Zwerger}},\ }\href
  {\doibase 10.1103/RevModPhys.80.885} {\bibfield  {journal} {\bibinfo
  {journal} {Rev. Mod. Phys.}\ }\textbf {\bibinfo {volume} {80}},\ \bibinfo
  {pages} {885} (\bibinfo {year} {2008})}\BibitemShut {NoStop}%
\bibitem [{\citenamefont {Lewenstein}\ \emph {et~al.}(2012)\citenamefont
  {Lewenstein}, \citenamefont {Sanpera},\ and\ \citenamefont
  {Ahufinger}}]{lewenstein2012ultracold}%
  \BibitemOpen
  \bibfield  {author} {\bibinfo {author} {\bibfnamefont {M.}~\bibnamefont
  {Lewenstein}}, \bibinfo {author} {\bibfnamefont {A.}~\bibnamefont {Sanpera}},
  \ and\ \bibinfo {author} {\bibfnamefont {V.}~\bibnamefont {Ahufinger}},\
  }\href@noop {} {\emph {\bibinfo {title} {Ultracold Atoms in Optical Lattices:
  Simulating quantum many-body systems}}}\ (\bibinfo  {publisher} {Oxford
  University Press},\ \bibinfo {year} {2012})\BibitemShut {NoStop}%
\bibitem [{\citenamefont {Bromley}\ \emph {et~al.}(2018)\citenamefont
  {Bromley}, \citenamefont {Kolkowitz}, \citenamefont {Bothwell}, \citenamefont
  {Kedar}, \citenamefont {Safavi-Naini}, \citenamefont {Wall}, \citenamefont
  {Salomon}, \citenamefont {Rey},\ and\ \citenamefont
  {Ye}}]{Bromley2018Dynamics}%
  \BibitemOpen
  \bibfield  {author} {\bibinfo {author} {\bibfnamefont {S.~L.}\ \bibnamefont
  {Bromley}}, \bibinfo {author} {\bibfnamefont {S.}~\bibnamefont {Kolkowitz}},
  \bibinfo {author} {\bibfnamefont {T.}~\bibnamefont {Bothwell}}, \bibinfo
  {author} {\bibfnamefont {D.}~\bibnamefont {Kedar}}, \bibinfo {author}
  {\bibfnamefont {A.}~\bibnamefont {Safavi-Naini}}, \bibinfo {author}
  {\bibfnamefont {M.~L.}\ \bibnamefont {Wall}}, \bibinfo {author}
  {\bibfnamefont {C.}~\bibnamefont {Salomon}}, \bibinfo {author} {\bibfnamefont
  {A.~M.}\ \bibnamefont {Rey}}, \ and\ \bibinfo {author} {\bibfnamefont
  {J.}~\bibnamefont {Ye}},\ }\href@noop {} {\bibfield  {journal} {\bibinfo
  {journal} {Nature Physics}\ }\textbf {\bibinfo {volume} {14}},\ \bibinfo
  {pages} {399} (\bibinfo {year} {2018})}\BibitemShut {NoStop}%
\bibitem [{\citenamefont {Goldman}\ \emph {et~al.}(2016)\citenamefont
  {Goldman}, \citenamefont {Budich},\ and\ \citenamefont
  {Zoller}}]{Goldman2016Topological}%
  \BibitemOpen
  \bibfield  {author} {\bibinfo {author} {\bibfnamefont {N.}~\bibnamefont
  {Goldman}}, \bibinfo {author} {\bibfnamefont {J.~C.}\ \bibnamefont {Budich}},
  \ and\ \bibinfo {author} {\bibfnamefont {P.}~\bibnamefont {Zoller}},\
  }\href@noop {} {\bibfield  {journal} {\bibinfo  {journal} {Nature Physics}\
  }\textbf {\bibinfo {volume} {12}},\ \bibinfo {pages} {639} (\bibinfo {year}
  {2016})}\BibitemShut {NoStop}%
\bibitem [{\citenamefont {Dalibard}\ \emph {et~al.}(2011)\citenamefont
  {Dalibard}, \citenamefont {Gerbier}, \citenamefont
  {Juzeli\ifmmode~\bar{u}\else \={u}\fi{}nas},\ and\ \citenamefont
  {\"Ohberg}}]{RevModPhys.83.1523}%
  \BibitemOpen
  \bibfield  {author} {\bibinfo {author} {\bibfnamefont {J.}~\bibnamefont
  {Dalibard}}, \bibinfo {author} {\bibfnamefont {F.}~\bibnamefont {Gerbier}},
  \bibinfo {author} {\bibfnamefont {G.}~\bibnamefont
  {Juzeli\ifmmode~\bar{u}\else \={u}\fi{}nas}}, \ and\ \bibinfo {author}
  {\bibfnamefont {P.}~\bibnamefont {\"Ohberg}},\ }\href {\doibase
  10.1103/RevModPhys.83.1523} {\bibfield  {journal} {\bibinfo  {journal} {Rev.
  Mod. Phys.}\ }\textbf {\bibinfo {volume} {83}},\ \bibinfo {pages} {1523}
  (\bibinfo {year} {2011})}\BibitemShut {NoStop}%
\bibitem [{\citenamefont {Goldman}\ \emph {et~al.}(2014)\citenamefont
  {Goldman}, \citenamefont {Juzeli{\=u}nas}, \citenamefont {{\"O}hberg},\ and\
  \citenamefont {Spielman}}]{goldman2014light}%
  \BibitemOpen
  \bibfield  {author} {\bibinfo {author} {\bibfnamefont {N.}~\bibnamefont
  {Goldman}}, \bibinfo {author} {\bibfnamefont {G.}~\bibnamefont
  {Juzeli{\=u}nas}}, \bibinfo {author} {\bibfnamefont {P.}~\bibnamefont
  {{\"O}hberg}}, \ and\ \bibinfo {author} {\bibfnamefont {I.~B.}\ \bibnamefont
  {Spielman}},\ }\href@noop {} {\bibfield  {journal} {\bibinfo  {journal}
  {Reports on Progress in Physics}\ }\textbf {\bibinfo {volume} {77}},\
  \bibinfo {pages} {126401} (\bibinfo {year} {2014})}\BibitemShut {NoStop}%
\bibitem [{\citenamefont {Cooper}\ \emph {et~al.}(2019)\citenamefont {Cooper},
  \citenamefont {Dalibard},\ and\ \citenamefont
  {Spielman}}]{RevModPhys.91.015005}%
  \BibitemOpen
  \bibfield  {author} {\bibinfo {author} {\bibfnamefont {N.~R.}\ \bibnamefont
  {Cooper}}, \bibinfo {author} {\bibfnamefont {J.}~\bibnamefont {Dalibard}}, \
  and\ \bibinfo {author} {\bibfnamefont {I.~B.}\ \bibnamefont {Spielman}},\
  }\href {\doibase 10.1103/RevModPhys.91.015005} {\bibfield  {journal}
  {\bibinfo  {journal} {Rev. Mod. Phys.}\ }\textbf {\bibinfo {volume} {91}},\
  \bibinfo {pages} {015005} (\bibinfo {year} {2019})}\BibitemShut {NoStop}%
\bibitem [{\citenamefont {Lin}\ \emph {et~al.}(2009)\citenamefont {Lin},
  \citenamefont {Compton}, \citenamefont {Jim\'{e}nez-Garc\'{i}a},
  \citenamefont {Porto},\ and\ \citenamefont {Spielman}}]{Y2009Synthetic}%
  \BibitemOpen
  \bibfield  {author} {\bibinfo {author} {\bibfnamefont {Y.-J.}\ \bibnamefont
  {Lin}}, \bibinfo {author} {\bibfnamefont {R.~L.}\ \bibnamefont {Compton}},
  \bibinfo {author} {\bibfnamefont {K.}~\bibnamefont {Jim\'{e}nez-Garc\'{i}a}},
  \bibinfo {author} {\bibfnamefont {J.~V.}\ \bibnamefont {Porto}}, \ and\
  \bibinfo {author} {\bibfnamefont {I.~B.}\ \bibnamefont {Spielman}},\
  }\href@noop {} {\bibfield  {journal} {\bibinfo  {journal} {Nature}\ }\textbf
  {\bibinfo {volume} {462}},\ \bibinfo {pages} {628} (\bibinfo {year}
  {2009})}\BibitemShut {NoStop}%
\bibitem [{\citenamefont {Lin}\ \emph {et~al.}(2010)\citenamefont {Lin},
  \citenamefont {Compton}, \citenamefont {Jim\'{e}nez-Garc\'{i}a},
  \citenamefont {Phillips}, \citenamefont {Porto},\ and\ \citenamefont
  {Spielman}}]{Lin2010A}%
  \BibitemOpen
  \bibfield  {author} {\bibinfo {author} {\bibfnamefont {Y.}~\bibnamefont
  {Lin}}, \bibinfo {author} {\bibfnamefont {R.~L.}\ \bibnamefont {Compton}},
  \bibinfo {author} {\bibfnamefont {K.}~\bibnamefont {Jim\'{e}nez-Garc\'{i}a}},
  \bibinfo {author} {\bibfnamefont {W.~D.}\ \bibnamefont {Phillips}}, \bibinfo
  {author} {\bibfnamefont {J.~V.}\ \bibnamefont {Porto}}, \ and\ \bibinfo
  {author} {\bibfnamefont {I.~B.}\ \bibnamefont {Spielman}},\ }\href@noop {}
  {\bibfield  {journal} {\bibinfo  {journal} {Nature Physics}\ }\textbf
  {\bibinfo {volume} {7}},\ \bibinfo {pages} {531} (\bibinfo {year}
  {2010})}\BibitemShut {NoStop}%
\bibitem [{\citenamefont {Mancini}\ \emph {et~al.}(2015)\citenamefont
  {Mancini}, \citenamefont {Pagano}, \citenamefont {Cappellini}, \citenamefont
  {Livi}, \citenamefont {Rider}, \citenamefont {Catani}, \citenamefont {Sias},
  \citenamefont {Zoller}, \citenamefont {Inguscio},\ and\ \citenamefont
  {Dalmonte}}]{Mancini2015Observation}%
  \BibitemOpen
  \bibfield  {author} {\bibinfo {author} {\bibfnamefont {M.}~\bibnamefont
  {Mancini}}, \bibinfo {author} {\bibfnamefont {G.}~\bibnamefont {Pagano}},
  \bibinfo {author} {\bibfnamefont {G.}~\bibnamefont {Cappellini}}, \bibinfo
  {author} {\bibfnamefont {L.}~\bibnamefont {Livi}}, \bibinfo {author}
  {\bibfnamefont {M.}~\bibnamefont {Rider}}, \bibinfo {author} {\bibfnamefont
  {J.}~\bibnamefont {Catani}}, \bibinfo {author} {\bibfnamefont
  {C.}~\bibnamefont {Sias}}, \bibinfo {author} {\bibfnamefont {P.}~\bibnamefont
  {Zoller}}, \bibinfo {author} {\bibfnamefont {M.}~\bibnamefont {Inguscio}}, \
  and\ \bibinfo {author} {\bibfnamefont {M.}~\bibnamefont {Dalmonte}},\
  }\href@noop {} {\bibfield  {journal} {\bibinfo  {journal} {Science}\ }\textbf
  {\bibinfo {volume} {349}},\ \bibinfo {pages} {1510} (\bibinfo {year}
  {2015})}\BibitemShut {NoStop}%
\bibitem [{\citenamefont {Livi}\ \emph {et~al.}(2016)\citenamefont {Livi},
  \citenamefont {Cappellini}, \citenamefont {Diem}, \citenamefont {Franchi},
  \citenamefont {Clivati}, \citenamefont {Frittelli}, \citenamefont {Levi},
  \citenamefont {Calonico}, \citenamefont {Catani}, \citenamefont {Inguscio},\
  and\ \citenamefont {Fallani}}]{Livi2016Synthetic}%
  \BibitemOpen
  \bibfield  {author} {\bibinfo {author} {\bibfnamefont {L.~F.}\ \bibnamefont
  {Livi}}, \bibinfo {author} {\bibfnamefont {G.}~\bibnamefont {Cappellini}},
  \bibinfo {author} {\bibfnamefont {M.}~\bibnamefont {Diem}}, \bibinfo {author}
  {\bibfnamefont {L.}~\bibnamefont {Franchi}}, \bibinfo {author} {\bibfnamefont
  {C.}~\bibnamefont {Clivati}}, \bibinfo {author} {\bibfnamefont
  {M.}~\bibnamefont {Frittelli}}, \bibinfo {author} {\bibfnamefont
  {F.}~\bibnamefont {Levi}}, \bibinfo {author} {\bibfnamefont {D.}~\bibnamefont
  {Calonico}}, \bibinfo {author} {\bibfnamefont {J.}~\bibnamefont {Catani}},
  \bibinfo {author} {\bibfnamefont {M.}~\bibnamefont {Inguscio}}, \ and\
  \bibinfo {author} {\bibfnamefont {L.}~\bibnamefont {Fallani}},\ }\href
  {\doibase 10.1103/PhysRevLett.117.220401} {\bibfield  {journal} {\bibinfo
  {journal} {Phys. Rev. Lett.}\ }\textbf {\bibinfo {volume} {117}},\ \bibinfo
  {pages} {220401} (\bibinfo {year} {2016})}\BibitemShut {NoStop}%
\bibitem [{\citenamefont {Atala}\ \emph {et~al.}(2014)\citenamefont {Atala},
  \citenamefont {Aidelsburger}, \citenamefont {Lohse}, \citenamefont
  {Barreiro}, \citenamefont {Paredes},\ and\ \citenamefont
  {Bloch}}]{Atala2014Observation}%
  \BibitemOpen
  \bibfield  {author} {\bibinfo {author} {\bibfnamefont {M.}~\bibnamefont
  {Atala}}, \bibinfo {author} {\bibfnamefont {M.}~\bibnamefont {Aidelsburger}},
  \bibinfo {author} {\bibfnamefont {M.}~\bibnamefont {Lohse}}, \bibinfo
  {author} {\bibfnamefont {J.}~\bibnamefont {Barreiro}}, \bibinfo {author}
  {\bibfnamefont {B.}~\bibnamefont {Paredes}}, \ and\ \bibinfo {author}
  {\bibfnamefont {I.}~\bibnamefont {Bloch}},\ }\href@noop {} {\bibfield
  {journal} {\bibinfo  {journal} {Nature Physics}\ }\textbf {\bibinfo {volume}
  {10}},\ \bibinfo {pages} {588} (\bibinfo {year} {2014})}\BibitemShut
  {NoStop}%
\bibitem [{\citenamefont {Stuhl}\ \emph {et~al.}(2015)\citenamefont {Stuhl},
  \citenamefont {Lu}, \citenamefont {Aycock}, \citenamefont {Genkina},\ and\
  \citenamefont {Spielman}}]{Stuhl2015Visualizing}%
  \BibitemOpen
  \bibfield  {author} {\bibinfo {author} {\bibfnamefont {B.~K.}\ \bibnamefont
  {Stuhl}}, \bibinfo {author} {\bibfnamefont {H.-I.}\ \bibnamefont {Lu}},
  \bibinfo {author} {\bibfnamefont {L.~M.}\ \bibnamefont {Aycock}}, \bibinfo
  {author} {\bibfnamefont {D.}~\bibnamefont {Genkina}}, \ and\ \bibinfo
  {author} {\bibfnamefont {I.~B.}\ \bibnamefont {Spielman}},\ }\href@noop {}
  {\bibfield  {journal} {\bibinfo  {journal} {Science}\ }\textbf {\bibinfo
  {volume} {349}},\ \bibinfo {pages} {1514} (\bibinfo {year}
  {2015})}\BibitemShut {NoStop}%
\bibitem [{\citenamefont {Kang}\ \emph {et~al.}(2018)\citenamefont {Kang},
  \citenamefont {Han},\ and\ \citenamefont {Shin}}]{PhysRevLett.121.150403}%
  \BibitemOpen
  \bibfield  {author} {\bibinfo {author} {\bibfnamefont {J.~H.}\ \bibnamefont
  {Kang}}, \bibinfo {author} {\bibfnamefont {J.~H.}\ \bibnamefont {Han}}, \
  and\ \bibinfo {author} {\bibfnamefont {Y.}~\bibnamefont {Shin}},\ }\href
  {\doibase 10.1103/PhysRevLett.121.150403} {\bibfield  {journal} {\bibinfo
  {journal} {Phys. Rev. Lett.}\ }\textbf {\bibinfo {volume} {121}},\ \bibinfo
  {pages} {150403} (\bibinfo {year} {2018})}\BibitemShut {NoStop}%
\bibitem [{\citenamefont {Aidelsburger}\ \emph {et~al.}(2011)\citenamefont
  {Aidelsburger}, \citenamefont {Atala}, \citenamefont {Nascimb\`{e}ne},
  \citenamefont {Trotzky}, \citenamefont {Chen},\ and\ \citenamefont
  {Bloch}}]{Aidelsburger2011Experimental}%
  \BibitemOpen
  \bibfield  {author} {\bibinfo {author} {\bibfnamefont {M.}~\bibnamefont
  {Aidelsburger}}, \bibinfo {author} {\bibfnamefont {M.}~\bibnamefont {Atala}},
  \bibinfo {author} {\bibfnamefont {S.}~\bibnamefont {Nascimb\`{e}ne}},
  \bibinfo {author} {\bibfnamefont {S.}~\bibnamefont {Trotzky}}, \bibinfo
  {author} {\bibfnamefont {Y.-A.}\ \bibnamefont {Chen}}, \ and\ \bibinfo
  {author} {\bibfnamefont {I.}~\bibnamefont {Bloch}},\ }\href@noop {}
  {\bibfield  {journal} {\bibinfo  {journal} {Physical Review Letters}\
  }\textbf {\bibinfo {volume} {107}},\ \bibinfo {pages} {487} (\bibinfo {year}
  {2011})}\BibitemShut {NoStop}%
\bibitem [{\citenamefont {Struck}\ \emph {et~al.}(2012)\citenamefont {Struck},
  \citenamefont {\"Olschl\"ager}, \citenamefont {Weinberg}, \citenamefont
  {Hauke}, \citenamefont {Simonet}, \citenamefont {Eckardt}, \citenamefont
  {Lewenstein}, \citenamefont {Sengstock},\ and\ \citenamefont
  {Windpassinger}}]{PhysRevLett.108.225304}%
  \BibitemOpen
  \bibfield  {author} {\bibinfo {author} {\bibfnamefont {J.}~\bibnamefont
  {Struck}}, \bibinfo {author} {\bibfnamefont {C.}~\bibnamefont
  {\"Olschl\"ager}}, \bibinfo {author} {\bibfnamefont {M.}~\bibnamefont
  {Weinberg}}, \bibinfo {author} {\bibfnamefont {P.}~\bibnamefont {Hauke}},
  \bibinfo {author} {\bibfnamefont {J.}~\bibnamefont {Simonet}}, \bibinfo
  {author} {\bibfnamefont {A.}~\bibnamefont {Eckardt}}, \bibinfo {author}
  {\bibfnamefont {M.}~\bibnamefont {Lewenstein}}, \bibinfo {author}
  {\bibfnamefont {K.}~\bibnamefont {Sengstock}}, \ and\ \bibinfo {author}
  {\bibfnamefont {P.}~\bibnamefont {Windpassinger}},\ }\href {\doibase
  10.1103/PhysRevLett.108.225304} {\bibfield  {journal} {\bibinfo  {journal}
  {Phys. Rev. Lett.}\ }\textbf {\bibinfo {volume} {108}},\ \bibinfo {pages}
  {225304} (\bibinfo {year} {2012})}\BibitemShut {NoStop}%
\bibitem [{\citenamefont {Aidelsburger}\ \emph {et~al.}(2013)\citenamefont
  {Aidelsburger}, \citenamefont {Atala}, \citenamefont {Lohse}, \citenamefont
  {Barreiro}, \citenamefont {Paredes},\ and\ \citenamefont
  {Bloch}}]{Aidelsburger2013Realization}%
  \BibitemOpen
  \bibfield  {author} {\bibinfo {author} {\bibfnamefont {M.}~\bibnamefont
  {Aidelsburger}}, \bibinfo {author} {\bibfnamefont {M.}~\bibnamefont {Atala}},
  \bibinfo {author} {\bibfnamefont {M.}~\bibnamefont {Lohse}}, \bibinfo
  {author} {\bibfnamefont {J.~T.}\ \bibnamefont {Barreiro}}, \bibinfo {author}
  {\bibfnamefont {B.}~\bibnamefont {Paredes}}, \ and\ \bibinfo {author}
  {\bibfnamefont {I.}~\bibnamefont {Bloch}},\ }\href {\doibase
  10.1103/PhysRevLett.111.185301} {\bibfield  {journal} {\bibinfo  {journal}
  {Phys. Rev. Lett.}\ }\textbf {\bibinfo {volume} {111}},\ \bibinfo {pages}
  {185301} (\bibinfo {year} {2013})}\BibitemShut {NoStop}%
\bibitem [{\citenamefont {Hirokazu}\ \emph {et~al.}(2013)\citenamefont
  {Hirokazu}, \citenamefont {Siviloglou}, \citenamefont {Kennedy},
  \citenamefont {William~Cody},\ and\ \citenamefont
  {Wolfgang}}]{Hirokazu2013Realizing}%
  \BibitemOpen
  \bibfield  {author} {\bibinfo {author} {\bibfnamefont {M.}~\bibnamefont
  {Hirokazu}}, \bibinfo {author} {\bibfnamefont {G.~A.}\ \bibnamefont
  {Siviloglou}}, \bibinfo {author} {\bibfnamefont {C.~J.}\ \bibnamefont
  {Kennedy}}, \bibinfo {author} {\bibfnamefont {B.}~\bibnamefont
  {William~Cody}}, \ and\ \bibinfo {author} {\bibfnamefont {K.}~\bibnamefont
  {Wolfgang}},\ }\href@noop {} {\bibfield  {journal} {\bibinfo  {journal}
  {Physical Review Letters}\ }\textbf {\bibinfo {volume} {111}},\ \bibinfo
  {pages} {185302} (\bibinfo {year} {2013})}\BibitemShut {NoStop}%
\bibitem [{\citenamefont {Gregor}\ \emph {et~al.}(2014)\citenamefont {Gregor},
  \citenamefont {Michael}, \citenamefont {R\'{e}mi}, \citenamefont {Martin},
  \citenamefont {Thomas}, \citenamefont {Daniel},\ and\ \citenamefont
  {Tilman}}]{Gregor2014Experimental}%
  \BibitemOpen
  \bibfield  {author} {\bibinfo {author} {\bibfnamefont {J.}~\bibnamefont
  {Gregor}}, \bibinfo {author} {\bibfnamefont {M.}~\bibnamefont {Michael}},
  \bibinfo {author} {\bibfnamefont {D.}~\bibnamefont {R\'{e}mi}}, \bibinfo
  {author} {\bibfnamefont {L.}~\bibnamefont {Martin}}, \bibinfo {author}
  {\bibfnamefont {U.}~\bibnamefont {Thomas}}, \bibinfo {author} {\bibfnamefont
  {G.}~\bibnamefont {Daniel}}, \ and\ \bibinfo {author} {\bibfnamefont
  {E.}~\bibnamefont {Tilman}},\ }\href@noop {} {\bibfield  {journal} {\bibinfo
  {journal} {Nature}\ }\textbf {\bibinfo {volume} {515}},\ \bibinfo {pages}
  {237} (\bibinfo {year} {2014})}\BibitemShut {NoStop}%
\bibitem [{\citenamefont {Kennedy}\ \emph {et~al.}(2015)\citenamefont
  {Kennedy}, \citenamefont {Burton}, \citenamefont {Chung},\ and\ \citenamefont
  {Ketterle}}]{Kennedy2015Observation}%
  \BibitemOpen
  \bibfield  {author} {\bibinfo {author} {\bibfnamefont {C.~J.}\ \bibnamefont
  {Kennedy}}, \bibinfo {author} {\bibfnamefont {W.~C.}\ \bibnamefont {Burton}},
  \bibinfo {author} {\bibfnamefont {W.~C.}\ \bibnamefont {Chung}}, \ and\
  \bibinfo {author} {\bibfnamefont {W.}~\bibnamefont {Ketterle}},\ }\href@noop
  {} {\bibfield  {journal} {\bibinfo  {journal} {Nature Physics}\ }\textbf
  {\bibinfo {volume} {11}},\ \bibinfo {pages} {1106} (\bibinfo {year}
  {2015})}\BibitemShut {NoStop}%
\bibitem [{\citenamefont {Fl{\"a}schner}\ \emph {et~al.}(2016)\citenamefont
  {Fl{\"a}schner}, \citenamefont {Rem}, \citenamefont {Tarnowski},
  \citenamefont {Vogel}, \citenamefont {L{\"u}hmann}, \citenamefont
  {Sengstock},\ and\ \citenamefont {Weitenberg}}]{Fl2016Experimental}%
  \BibitemOpen
  \bibfield  {author} {\bibinfo {author} {\bibfnamefont {N.}~\bibnamefont
  {Fl{\"a}schner}}, \bibinfo {author} {\bibfnamefont {B.~S.}\ \bibnamefont
  {Rem}}, \bibinfo {author} {\bibfnamefont {M.}~\bibnamefont {Tarnowski}},
  \bibinfo {author} {\bibfnamefont {D.}~\bibnamefont {Vogel}}, \bibinfo
  {author} {\bibfnamefont {D.-S.}\ \bibnamefont {L{\"u}hmann}}, \bibinfo
  {author} {\bibfnamefont {K.}~\bibnamefont {Sengstock}}, \ and\ \bibinfo
  {author} {\bibfnamefont {C.}~\bibnamefont {Weitenberg}},\ }\href {\doibase
  10.1126/science.aad4568} {\bibfield  {journal} {\bibinfo  {journal}
  {Science}\ }\textbf {\bibinfo {volume} {352}},\ \bibinfo {pages} {1091}
  (\bibinfo {year} {2016})}\BibitemShut {NoStop}%
\bibitem [{\citenamefont {Asteria}\ \emph {et~al.}(2019)\citenamefont
  {Asteria}, \citenamefont {Tran}, \citenamefont {Ozawa}, \citenamefont
  {Tarnowski}, \citenamefont {Rem}, \citenamefont {Fl{\"a}schner},
  \citenamefont {Sengstock}, \citenamefont {Goldman},\ and\ \citenamefont
  {Weitenberg}}]{Asteria2018Measuring}%
  \BibitemOpen
  \bibfield  {author} {\bibinfo {author} {\bibfnamefont {L.}~\bibnamefont
  {Asteria}}, \bibinfo {author} {\bibfnamefont {D.~T.}\ \bibnamefont {Tran}},
  \bibinfo {author} {\bibfnamefont {T.}~\bibnamefont {Ozawa}}, \bibinfo
  {author} {\bibfnamefont {M.}~\bibnamefont {Tarnowski}}, \bibinfo {author}
  {\bibfnamefont {B.~S.}\ \bibnamefont {Rem}}, \bibinfo {author} {\bibfnamefont
  {N.}~\bibnamefont {Fl{\"a}schner}}, \bibinfo {author} {\bibfnamefont
  {K.}~\bibnamefont {Sengstock}}, \bibinfo {author} {\bibfnamefont
  {B.}~\bibnamefont {Goldman}}, \ and\ \bibinfo {author} {\bibfnamefont
  {C.}~\bibnamefont {Weitenberg}},\ }\href@noop {} {\bibfield  {journal}
  {\bibinfo  {journal} {Nature Physics}\ }\textbf {\bibinfo {volume} {15}},\
  \bibinfo {pages} {449} (\bibinfo {year} {2019})}\BibitemShut {NoStop}%
\bibitem [{\citenamefont {Wall}\ \emph {et~al.}(2016)\citenamefont {Wall},
  \citenamefont {Koller}, \citenamefont {Li}, \citenamefont {Zhang},
  \citenamefont {Cooper}, \citenamefont {Ye},\ and\ \citenamefont
  {Rey}}]{Wall2016Synthetic}%
  \BibitemOpen
  \bibfield  {author} {\bibinfo {author} {\bibfnamefont {M.~L.}\ \bibnamefont
  {Wall}}, \bibinfo {author} {\bibfnamefont {A.~P.}\ \bibnamefont {Koller}},
  \bibinfo {author} {\bibfnamefont {S.}~\bibnamefont {Li}}, \bibinfo {author}
  {\bibfnamefont {X.}~\bibnamefont {Zhang}}, \bibinfo {author} {\bibfnamefont
  {N.~R.}\ \bibnamefont {Cooper}}, \bibinfo {author} {\bibfnamefont
  {J.}~\bibnamefont {Ye}}, \ and\ \bibinfo {author} {\bibfnamefont {A.~M.}\
  \bibnamefont {Rey}},\ }\href@noop {} {\bibfield  {journal} {\bibinfo
  {journal} {Physical Review Letters}\ }\textbf {\bibinfo {volume} {116}},\
  \bibinfo {pages} {035301} (\bibinfo {year} {2016})}\BibitemShut {NoStop}%
\bibitem [{\citenamefont {Zhou}\ \emph {et~al.}(2017)\citenamefont {Zhou},
  \citenamefont {Pan}, \citenamefont {Liu}, \citenamefont {Zhang},
  \citenamefont {Yi}, \citenamefont {Chen},\ and\ \citenamefont
  {Jia}}]{PhysRevLett.119.185701}%
  \BibitemOpen
  \bibfield  {author} {\bibinfo {author} {\bibfnamefont {X.}~\bibnamefont
  {Zhou}}, \bibinfo {author} {\bibfnamefont {J.-S.}\ \bibnamefont {Pan}},
  \bibinfo {author} {\bibfnamefont {Z.-X.}\ \bibnamefont {Liu}}, \bibinfo
  {author} {\bibfnamefont {W.}~\bibnamefont {Zhang}}, \bibinfo {author}
  {\bibfnamefont {W.}~\bibnamefont {Yi}}, \bibinfo {author} {\bibfnamefont
  {G.}~\bibnamefont {Chen}}, \ and\ \bibinfo {author} {\bibfnamefont
  {S.}~\bibnamefont {Jia}},\ }\href {\doibase 10.1103/PhysRevLett.119.185701}
  {\bibfield  {journal} {\bibinfo  {journal} {Phys. Rev. Lett.}\ }\textbf
  {\bibinfo {volume} {119}},\ \bibinfo {pages} {185701} (\bibinfo {year}
  {2017})}\BibitemShut {NoStop}%
\bibitem [{\citenamefont {Kolkowitz}\ \emph {et~al.}(2016)\citenamefont
  {Kolkowitz}, \citenamefont {Bromley}, \citenamefont {Bothwell}, \citenamefont
  {Wall}, \citenamefont {Marti}, \citenamefont {Koller}, \citenamefont {Zhang},
  \citenamefont {Rey},\ and\ \citenamefont {Ye}}]{Kolkowitz2016Spin}%
  \BibitemOpen
  \bibfield  {author} {\bibinfo {author} {\bibfnamefont {S.}~\bibnamefont
  {Kolkowitz}}, \bibinfo {author} {\bibfnamefont {S.~L.}\ \bibnamefont
  {Bromley}}, \bibinfo {author} {\bibfnamefont {T.}~\bibnamefont {Bothwell}},
  \bibinfo {author} {\bibfnamefont {M.~L.}\ \bibnamefont {Wall}}, \bibinfo
  {author} {\bibfnamefont {G.~E.}\ \bibnamefont {Marti}}, \bibinfo {author}
  {\bibfnamefont {A.~P.}\ \bibnamefont {Koller}}, \bibinfo {author}
  {\bibfnamefont {X.}~\bibnamefont {Zhang}}, \bibinfo {author} {\bibfnamefont
  {A.~M.}\ \bibnamefont {Rey}}, \ and\ \bibinfo {author} {\bibfnamefont
  {J.}~\bibnamefont {Ye}},\ }\href@noop {} {\bibfield  {journal} {\bibinfo
  {journal} {Nature}\ }\textbf {\bibinfo {volume} {542}},\ \bibinfo {pages}
  {66} (\bibinfo {year} {2016})}\BibitemShut {NoStop}%
\bibitem [{\citenamefont {Tai}\ \emph {et~al.}(2017)\citenamefont {Tai},
  \citenamefont {Lukin}, \citenamefont {Rispoli}, \citenamefont {Schittko},
  \citenamefont {Menke}, \citenamefont {Dan}, \citenamefont {Preiss},
  \citenamefont {Grusdt}, \citenamefont {Kaufman},\ and\ \citenamefont
  {Greiner}}]{Tai2017Microscopy}%
  \BibitemOpen
  \bibfield  {author} {\bibinfo {author} {\bibfnamefont {M.~E.}\ \bibnamefont
  {Tai}}, \bibinfo {author} {\bibfnamefont {A.}~\bibnamefont {Lukin}}, \bibinfo
  {author} {\bibfnamefont {M.}~\bibnamefont {Rispoli}}, \bibinfo {author}
  {\bibfnamefont {R.}~\bibnamefont {Schittko}}, \bibinfo {author}
  {\bibfnamefont {T.}~\bibnamefont {Menke}}, \bibinfo {author} {\bibfnamefont
  {B.}~\bibnamefont {Dan}}, \bibinfo {author} {\bibfnamefont {P.~M.}\
  \bibnamefont {Preiss}}, \bibinfo {author} {\bibfnamefont {F.}~\bibnamefont
  {Grusdt}}, \bibinfo {author} {\bibfnamefont {A.~M.}\ \bibnamefont {Kaufman}},
  \ and\ \bibinfo {author} {\bibfnamefont {M.}~\bibnamefont {Greiner}},\
  }\href@noop {} {\bibfield  {journal} {\bibinfo  {journal} {Nature}\ }\textbf
  {\bibinfo {volume} {546}},\ \bibinfo {pages} {519} (\bibinfo {year}
  {2017})}\BibitemShut {NoStop}%
\bibitem [{\citenamefont {He}\ \emph {et~al.}(2019)\citenamefont {He},
  \citenamefont {Perlin}, \citenamefont {Muleady}, \citenamefont {Lewis-Swan},
  \citenamefont {Hutson}, \citenamefont {Ye},\ and\ \citenamefont
  {Rey}}]{Rey_2019}%
  \BibitemOpen
  \bibfield  {author} {\bibinfo {author} {\bibfnamefont {P.}~\bibnamefont
  {He}}, \bibinfo {author} {\bibfnamefont {M.~A.}\ \bibnamefont {Perlin}},
  \bibinfo {author} {\bibfnamefont {S.~R.}\ \bibnamefont {Muleady}}, \bibinfo
  {author} {\bibfnamefont {R.~J.}\ \bibnamefont {Lewis-Swan}}, \bibinfo
  {author} {\bibfnamefont {R.~B.}\ \bibnamefont {Hutson}}, \bibinfo {author}
  {\bibfnamefont {J.}~\bibnamefont {Ye}}, \ and\ \bibinfo {author}
  {\bibfnamefont {A.~M.}\ \bibnamefont {Rey}},\ }\href@noop {} {\bibfield
  {journal} {\bibinfo  {journal} {1904.07866}\ } (\bibinfo {year}
  {2019})}\BibitemShut {NoStop}%
\bibitem [{\citenamefont {Saffman}\ \emph {et~al.}(2010)\citenamefont
  {Saffman}, \citenamefont {Walker},\ and\ \citenamefont
  {M\o{}lmer}}]{RevModPhys.82.2313}%
  \BibitemOpen
  \bibfield  {author} {\bibinfo {author} {\bibfnamefont {M.}~\bibnamefont
  {Saffman}}, \bibinfo {author} {\bibfnamefont {T.~G.}\ \bibnamefont {Walker}},
  \ and\ \bibinfo {author} {\bibfnamefont {K.}~\bibnamefont {M\o{}lmer}},\
  }\href {\doibase 10.1103/RevModPhys.82.2313} {\bibfield  {journal} {\bibinfo
  {journal} {Rev. Mod. Phys.}\ }\textbf {\bibinfo {volume} {82}},\ \bibinfo
  {pages} {2313} (\bibinfo {year} {2010})}\BibitemShut {NoStop}%
\bibitem [{\citenamefont {Labuhn}\ \emph {et~al.}(2016)\citenamefont {Labuhn},
  \citenamefont {Barredo}, \citenamefont {Ravets}, \citenamefont
  {De~L{\'e}s{\'e}leuc}, \citenamefont {Macr{\`\i}}, \citenamefont {Lahaye},\
  and\ \citenamefont {Browaeys}}]{labuhn2016tunable}%
  \BibitemOpen
  \bibfield  {author} {\bibinfo {author} {\bibfnamefont {H.}~\bibnamefont
  {Labuhn}}, \bibinfo {author} {\bibfnamefont {D.}~\bibnamefont {Barredo}},
  \bibinfo {author} {\bibfnamefont {S.}~\bibnamefont {Ravets}}, \bibinfo
  {author} {\bibfnamefont {S.}~\bibnamefont {De~L{\'e}s{\'e}leuc}}, \bibinfo
  {author} {\bibfnamefont {T.}~\bibnamefont {Macr{\`\i}}}, \bibinfo {author}
  {\bibfnamefont {T.}~\bibnamefont {Lahaye}}, \ and\ \bibinfo {author}
  {\bibfnamefont {A.}~\bibnamefont {Browaeys}},\ }\href@noop {} {\bibfield
  {journal} {\bibinfo  {journal} {Nature}\ }\textbf {\bibinfo {volume} {534}},\
  \bibinfo {pages} {667} (\bibinfo {year} {2016})}\BibitemShut {NoStop}%
\bibitem [{\citenamefont {Levine}\ \emph {et~al.}(2018)\citenamefont {Levine},
  \citenamefont {Keesling}, \citenamefont {Omran}, \citenamefont {Bernien},
  \citenamefont {Schwartz}, \citenamefont {Zibrov}, \citenamefont {Endres},
  \citenamefont {Greiner}, \citenamefont {Vuleti\ifmmode~\acute{c}\else
  \'{c}\fi{}},\ and\ \citenamefont {Lukin}}]{PhysRevLett.121.123603}%
  \BibitemOpen
  \bibfield  {author} {\bibinfo {author} {\bibfnamefont {H.}~\bibnamefont
  {Levine}}, \bibinfo {author} {\bibfnamefont {A.}~\bibnamefont {Keesling}},
  \bibinfo {author} {\bibfnamefont {A.}~\bibnamefont {Omran}}, \bibinfo
  {author} {\bibfnamefont {H.}~\bibnamefont {Bernien}}, \bibinfo {author}
  {\bibfnamefont {S.}~\bibnamefont {Schwartz}}, \bibinfo {author}
  {\bibfnamefont {A.~S.}\ \bibnamefont {Zibrov}}, \bibinfo {author}
  {\bibfnamefont {M.}~\bibnamefont {Endres}}, \bibinfo {author} {\bibfnamefont
  {M.}~\bibnamefont {Greiner}}, \bibinfo {author} {\bibfnamefont
  {V.}~\bibnamefont {Vuleti\ifmmode~\acute{c}\else \'{c}\fi{}}}, \ and\
  \bibinfo {author} {\bibfnamefont {M.~D.}\ \bibnamefont {Lukin}},\ }\href
  {\doibase 10.1103/PhysRevLett.121.123603} {\bibfield  {journal} {\bibinfo
  {journal} {Phys. Rev. Lett.}\ }\textbf {\bibinfo {volume} {121}},\ \bibinfo
  {pages} {123603} (\bibinfo {year} {2018})}\BibitemShut {NoStop}%
\bibitem [{\citenamefont {Omran}\ \emph {et~al.}(2019)\citenamefont {Omran},
  \citenamefont {Levine}, \citenamefont {Keesling}, \citenamefont {Semeghini},
  \citenamefont {Wang}, \citenamefont {Ebadi}, \citenamefont {Bernien},
  \citenamefont {Zibrov}, \citenamefont {Pichler}, \citenamefont {Choi} \emph
  {et~al.}}]{omran2019generation}%
  \BibitemOpen
  \bibfield  {author} {\bibinfo {author} {\bibfnamefont {A.}~\bibnamefont
  {Omran}}, \bibinfo {author} {\bibfnamefont {H.}~\bibnamefont {Levine}},
  \bibinfo {author} {\bibfnamefont {A.}~\bibnamefont {Keesling}}, \bibinfo
  {author} {\bibfnamefont {G.}~\bibnamefont {Semeghini}}, \bibinfo {author}
  {\bibfnamefont {T.~T.}\ \bibnamefont {Wang}}, \bibinfo {author}
  {\bibfnamefont {S.}~\bibnamefont {Ebadi}}, \bibinfo {author} {\bibfnamefont
  {H.}~\bibnamefont {Bernien}}, \bibinfo {author} {\bibfnamefont {A.~S.}\
  \bibnamefont {Zibrov}}, \bibinfo {author} {\bibfnamefont {H.}~\bibnamefont
  {Pichler}}, \bibinfo {author} {\bibfnamefont {S.}~\bibnamefont {Choi}},
  \emph {et~al.},\ }\href@noop {} {\bibfield  {journal} {\bibinfo  {journal}
  {Science}\ }\textbf {\bibinfo {volume} {365}},\ \bibinfo {pages} {570}
  (\bibinfo {year} {2019})}\BibitemShut {NoStop}%
\bibitem [{\citenamefont {de~L{\'e}s{\'e}leuc}\ \emph
  {et~al.}(2019)\citenamefont {de~L{\'e}s{\'e}leuc}, \citenamefont {Lienhard},
  \citenamefont {Scholl}, \citenamefont {Barredo}, \citenamefont {Weber},
  \citenamefont {Lang}, \citenamefont {B{\"u}chler}, \citenamefont {Lahaye},\
  and\ \citenamefont {Browaeys}}]{de2018experimental}%
  \BibitemOpen
  \bibfield  {author} {\bibinfo {author} {\bibfnamefont {S.}~\bibnamefont
  {de~L{\'e}s{\'e}leuc}}, \bibinfo {author} {\bibfnamefont {V.}~\bibnamefont
  {Lienhard}}, \bibinfo {author} {\bibfnamefont {P.}~\bibnamefont {Scholl}},
  \bibinfo {author} {\bibfnamefont {D.}~\bibnamefont {Barredo}}, \bibinfo
  {author} {\bibfnamefont {S.}~\bibnamefont {Weber}}, \bibinfo {author}
  {\bibfnamefont {N.}~\bibnamefont {Lang}}, \bibinfo {author} {\bibfnamefont
  {H.~P.}\ \bibnamefont {B{\"u}chler}}, \bibinfo {author} {\bibfnamefont
  {T.}~\bibnamefont {Lahaye}}, \ and\ \bibinfo {author} {\bibfnamefont
  {A.}~\bibnamefont {Browaeys}},\ }\href@noop {} {\bibfield  {journal}
  {\bibinfo  {journal} {Science}\ }\textbf {\bibinfo {volume} {365}},\ \bibinfo
  {pages} {775} (\bibinfo {year} {2019})}\BibitemShut {NoStop}%
\bibitem [{\citenamefont {Celi}\ \emph {et~al.}(2019)\citenamefont {Celi},
  \citenamefont {Vermersch}, \citenamefont {Viyuela}, \citenamefont {Pichler},
  \citenamefont {Lukin},\ and\ \citenamefont {Zoller}}]{celi2019emerging}%
  \BibitemOpen
  \bibfield  {author} {\bibinfo {author} {\bibfnamefont {A.}~\bibnamefont
  {Celi}}, \bibinfo {author} {\bibfnamefont {B.}~\bibnamefont {Vermersch}},
  \bibinfo {author} {\bibfnamefont {O.}~\bibnamefont {Viyuela}}, \bibinfo
  {author} {\bibfnamefont {H.}~\bibnamefont {Pichler}}, \bibinfo {author}
  {\bibfnamefont {M.~D.}\ \bibnamefont {Lukin}}, \ and\ \bibinfo {author}
  {\bibfnamefont {P.}~\bibnamefont {Zoller}},\ }\href@noop {} {\bibfield
  {journal} {\bibinfo  {journal} {arXiv preprint arXiv:1907.03311}\ } (\bibinfo
  {year} {2019})}\BibitemShut {NoStop}%
\bibitem [{\citenamefont {Bouchoule}\ and\ \citenamefont
  {M\o{}lmer}(2002)}]{PhysRevA.65.041803}%
  \BibitemOpen
  \bibfield  {author} {\bibinfo {author} {\bibfnamefont {I.}~\bibnamefont
  {Bouchoule}}\ and\ \bibinfo {author} {\bibfnamefont {K.}~\bibnamefont
  {M\o{}lmer}},\ }\href {\doibase 10.1103/PhysRevA.65.041803} {\bibfield
  {journal} {\bibinfo  {journal} {Phys. Rev. A}\ }\textbf {\bibinfo {volume}
  {65}},\ \bibinfo {pages} {041803} (\bibinfo {year} {2002})}\BibitemShut
  {NoStop}%
\bibitem [{\citenamefont {Pupillo}\ \emph {et~al.}(2010)\citenamefont
  {Pupillo}, \citenamefont {Micheli}, \citenamefont {Boninsegni}, \citenamefont
  {Lesanovsky},\ and\ \citenamefont {Zoller}}]{PhysRevLett.104.223002}%
  \BibitemOpen
  \bibfield  {author} {\bibinfo {author} {\bibfnamefont {G.}~\bibnamefont
  {Pupillo}}, \bibinfo {author} {\bibfnamefont {A.}~\bibnamefont {Micheli}},
  \bibinfo {author} {\bibfnamefont {M.}~\bibnamefont {Boninsegni}}, \bibinfo
  {author} {\bibfnamefont {I.}~\bibnamefont {Lesanovsky}}, \ and\ \bibinfo
  {author} {\bibfnamefont {P.}~\bibnamefont {Zoller}},\ }\href {\doibase
  10.1103/PhysRevLett.104.223002} {\bibfield  {journal} {\bibinfo  {journal}
  {Phys. Rev. Lett.}\ }\textbf {\bibinfo {volume} {104}},\ \bibinfo {pages}
  {223002} (\bibinfo {year} {2010})}\BibitemShut {NoStop}%
\bibitem [{\citenamefont {Honer}\ \emph {et~al.}(2010)\citenamefont {Honer},
  \citenamefont {Weimer}, \citenamefont {Pfau},\ and\ \citenamefont
  {B\"uchler}}]{Honer_10}%
  \BibitemOpen
  \bibfield  {author} {\bibinfo {author} {\bibfnamefont {J.}~\bibnamefont
  {Honer}}, \bibinfo {author} {\bibfnamefont {H.}~\bibnamefont {Weimer}},
  \bibinfo {author} {\bibfnamefont {T.}~\bibnamefont {Pfau}}, \ and\ \bibinfo
  {author} {\bibfnamefont {H.~P.}\ \bibnamefont {B\"uchler}},\ }\href {\doibase
  10.1103/PhysRevLett.105.160404} {\bibfield  {journal} {\bibinfo  {journal}
  {Phys. Rev. Lett.}\ }\textbf {\bibinfo {volume} {105}},\ \bibinfo {pages}
  {160404} (\bibinfo {year} {2010})}\BibitemShut {NoStop}%
\bibitem [{\citenamefont {Cinti}\ \emph {et~al.}(2010)\citenamefont {Cinti},
  \citenamefont {Jain}, \citenamefont {Boninsegni}, \citenamefont {Micheli},
  \citenamefont {Zoller},\ and\ \citenamefont {Pupillo}}]{Cinti_10}%
  \BibitemOpen
  \bibfield  {author} {\bibinfo {author} {\bibfnamefont {F.}~\bibnamefont
  {Cinti}}, \bibinfo {author} {\bibfnamefont {P.}~\bibnamefont {Jain}},
  \bibinfo {author} {\bibfnamefont {M.}~\bibnamefont {Boninsegni}}, \bibinfo
  {author} {\bibfnamefont {A.}~\bibnamefont {Micheli}}, \bibinfo {author}
  {\bibfnamefont {P.}~\bibnamefont {Zoller}}, \ and\ \bibinfo {author}
  {\bibfnamefont {G.}~\bibnamefont {Pupillo}},\ }\href {\doibase
  10.1103/PhysRevLett.105.135301} {\bibfield  {journal} {\bibinfo  {journal}
  {Phys. Rev. Lett.}\ }\textbf {\bibinfo {volume} {105}},\ \bibinfo {pages}
  {135301} (\bibinfo {year} {2010})}\BibitemShut {NoStop}%
\bibitem [{\citenamefont {Li}\ \emph {et~al.}(2012)\citenamefont {Li},
  \citenamefont {Hamadeh},\ and\ \citenamefont {Lesanovsky}}]{Li_12}%
  \BibitemOpen
  \bibfield  {author} {\bibinfo {author} {\bibfnamefont {W.}~\bibnamefont
  {Li}}, \bibinfo {author} {\bibfnamefont {L.}~\bibnamefont {Hamadeh}}, \ and\
  \bibinfo {author} {\bibfnamefont {I.}~\bibnamefont {Lesanovsky}},\ }\href
  {\doibase 10.1103/PhysRevA.85.053615} {\bibfield  {journal} {\bibinfo
  {journal} {Phys. Rev. A}\ }\textbf {\bibinfo {volume} {85}},\ \bibinfo
  {pages} {053615} (\bibinfo {year} {2012})}\BibitemShut {NoStop}%
\bibitem [{\citenamefont {Henkel}\ \emph {et~al.}(2010)\citenamefont {Henkel},
  \citenamefont {Nath},\ and\ \citenamefont {Pohl}}]{Henkel_10}%
  \BibitemOpen
  \bibfield  {author} {\bibinfo {author} {\bibfnamefont {N.}~\bibnamefont
  {Henkel}}, \bibinfo {author} {\bibfnamefont {R.}~\bibnamefont {Nath}}, \ and\
  \bibinfo {author} {\bibfnamefont {T.}~\bibnamefont {Pohl}},\ }\href {\doibase
  10.1103/PhysRevLett.104.195302} {\bibfield  {journal} {\bibinfo  {journal}
  {Phys. Rev. Lett.}\ }\textbf {\bibinfo {volume} {104}},\ \bibinfo {pages}
  {195302} (\bibinfo {year} {2010})}\BibitemShut {NoStop}%
\bibitem [{\citenamefont {Bounds}\ \emph {et~al.}(2018)\citenamefont {Bounds},
  \citenamefont {Jackson}, \citenamefont {Hanley}, \citenamefont {Faoro},
  \citenamefont {Bridge}, \citenamefont {Huillery},\ and\ \citenamefont
  {Jones}}]{dress_trap}%
  \BibitemOpen
  \bibfield  {author} {\bibinfo {author} {\bibfnamefont {A.~D.}\ \bibnamefont
  {Bounds}}, \bibinfo {author} {\bibfnamefont {N.~C.}\ \bibnamefont {Jackson}},
  \bibinfo {author} {\bibfnamefont {R.~K.}\ \bibnamefont {Hanley}}, \bibinfo
  {author} {\bibfnamefont {R.}~\bibnamefont {Faoro}}, \bibinfo {author}
  {\bibfnamefont {E.~M.}\ \bibnamefont {Bridge}}, \bibinfo {author}
  {\bibfnamefont {P.}~\bibnamefont {Huillery}}, \ and\ \bibinfo {author}
  {\bibfnamefont {M.~P.~A.}\ \bibnamefont {Jones}},\ }\href {\doibase
  10.1103/PhysRevLett.120.183401} {\bibfield  {journal} {\bibinfo  {journal}
  {Phys. Rev. Lett.}\ }\textbf {\bibinfo {volume} {120}},\ \bibinfo {pages}
  {183401} (\bibinfo {year} {2018})}\BibitemShut {NoStop}%
\bibitem [{\citenamefont {Jau}\ \emph {et~al.}(2016)\citenamefont {Jau},
  \citenamefont {Hankin}, \citenamefont {Keating}, \citenamefont {Deutsch},\
  and\ \citenamefont {Biedermann}}]{jau_entangling_2016}%
  \BibitemOpen
  \bibfield  {author} {\bibinfo {author} {\bibfnamefont {Y.-Y.}\ \bibnamefont
  {Jau}}, \bibinfo {author} {\bibfnamefont {A.~M.}\ \bibnamefont {Hankin}},
  \bibinfo {author} {\bibfnamefont {T.}~\bibnamefont {Keating}}, \bibinfo
  {author} {\bibfnamefont {I.~H.}\ \bibnamefont {Deutsch}}, \ and\ \bibinfo
  {author} {\bibfnamefont {G.~W.}\ \bibnamefont {Biedermann}},\ }\href
  {\doibase 10.1038/nphys3487} {\bibfield  {journal} {\bibinfo  {journal}
  {Nature Physics}\ }\textbf {\bibinfo {volume} {12}},\ \bibinfo {pages} {71}
  (\bibinfo {year} {2016})}\BibitemShut {NoStop}%
\bibitem [{\citenamefont {Zeiher}\ \emph {et~al.}(2016)\citenamefont {Zeiher},
  \citenamefont {Van~Bijnen}, \citenamefont {Schau{\ss}}, \citenamefont {Hild},
  \citenamefont {Choi}, \citenamefont {Pohl}, \citenamefont {Bloch},\ and\
  \citenamefont {Gross}}]{zeiher2016many}%
  \BibitemOpen
  \bibfield  {author} {\bibinfo {author} {\bibfnamefont {J.}~\bibnamefont
  {Zeiher}}, \bibinfo {author} {\bibfnamefont {R.}~\bibnamefont {Van~Bijnen}},
  \bibinfo {author} {\bibfnamefont {P.}~\bibnamefont {Schau{\ss}}}, \bibinfo
  {author} {\bibfnamefont {S.}~\bibnamefont {Hild}}, \bibinfo {author}
  {\bibfnamefont {J.-y.}\ \bibnamefont {Choi}}, \bibinfo {author}
  {\bibfnamefont {T.}~\bibnamefont {Pohl}}, \bibinfo {author} {\bibfnamefont
  {I.}~\bibnamefont {Bloch}}, \ and\ \bibinfo {author} {\bibfnamefont
  {C.}~\bibnamefont {Gross}},\ }\href@noop {} {\bibfield  {journal} {\bibinfo
  {journal} {Nature Physics}\ }\textbf {\bibinfo {volume} {12}},\ \bibinfo
  {pages} {1095} (\bibinfo {year} {2016})}\BibitemShut {NoStop}%
\bibitem [{\citenamefont {Wang}\ \emph {et~al.}(2015)\citenamefont {Wang},
  \citenamefont {Liu}, \citenamefont {Zhu},\ and\ \citenamefont
  {Scully}}]{wang_superradiance_2015}%
  \BibitemOpen
  \bibfield  {author} {\bibinfo {author} {\bibfnamefont {D.-W.}\ \bibnamefont
  {Wang}}, \bibinfo {author} {\bibfnamefont {R.-B.}\ \bibnamefont {Liu}},
  \bibinfo {author} {\bibfnamefont {S.-Y.}\ \bibnamefont {Zhu}}, \ and\
  \bibinfo {author} {\bibfnamefont {M.~O.}\ \bibnamefont {Scully}},\ }\href
  {\doibase 10.1103/PhysRevLett.114.043602} {\bibfield  {journal} {\bibinfo
  {journal} {Phys. Rev. Lett.}\ }\textbf {\bibinfo {volume} {114}},\ \bibinfo
  {pages} {043602} (\bibinfo {year} {2015})}\BibitemShut {NoStop}%
\bibitem [{\citenamefont {Cai}\ \emph {et~al.}(2019)\citenamefont {Cai},
  \citenamefont {Liu}, \citenamefont {Wu}, \citenamefont {He}, \citenamefont
  {Zhu}, \citenamefont {Zhang},\ and\ \citenamefont
  {Wang}}]{PhysRevLett.122.023601}%
  \BibitemOpen
  \bibfield  {author} {\bibinfo {author} {\bibfnamefont {H.}~\bibnamefont
  {Cai}}, \bibinfo {author} {\bibfnamefont {J.}~\bibnamefont {Liu}}, \bibinfo
  {author} {\bibfnamefont {J.}~\bibnamefont {Wu}}, \bibinfo {author}
  {\bibfnamefont {Y.}~\bibnamefont {He}}, \bibinfo {author} {\bibfnamefont
  {S.-Y.}\ \bibnamefont {Zhu}}, \bibinfo {author} {\bibfnamefont {J.-X.}\
  \bibnamefont {Zhang}}, \ and\ \bibinfo {author} {\bibfnamefont {D.-W.}\
  \bibnamefont {Wang}},\ }\href {\doibase 10.1103/PhysRevLett.122.023601}
  {\bibfield  {journal} {\bibinfo  {journal} {Phys. Rev. Lett.}\ }\textbf
  {\bibinfo {volume} {122}},\ \bibinfo {pages} {023601} (\bibinfo {year}
  {2019})}\BibitemShut {NoStop}%
\bibitem [{exp()}]{explain_2}%
  \BibitemOpen
  \href@noop {} {\bibinfo  {journal} {We consider the system with lattice sites
  up to $L_{\rm lat}=256$ to verify the finite-size effects on phase diagrams}\
  }\BibitemShut {NoStop}%
\bibitem [{\citenamefont {Vasi{\'c}}\ \emph {et~al.}(2015)\citenamefont
  {Vasi{\'c}}, \citenamefont {Petrescu}, \citenamefont {Le~Hur},\ and\
  \citenamefont {Hofstetter}}]{vasic2015chiral}%
  \BibitemOpen
\bibfield  {journal} {  }\bibfield  {author} {\bibinfo {author} {\bibfnamefont
  {I.}~\bibnamefont {Vasi{\'c}}}, \bibinfo {author} {\bibfnamefont
  {A.}~\bibnamefont {Petrescu}}, \bibinfo {author} {\bibfnamefont
  {K.}~\bibnamefont {Le~Hur}}, \ and\ \bibinfo {author} {\bibfnamefont
  {W.}~\bibnamefont {Hofstetter}},\ }\href@noop {} {\bibfield  {journal}
  {\bibinfo  {journal} {Phys. Rev. B}\ }\textbf {\bibinfo {volume} {91}},\
  \bibinfo {pages} {094502} (\bibinfo {year} {2015})}\BibitemShut {NoStop}%
\bibitem [{\citenamefont {Plekhanov}\ \emph {et~al.}(2018)\citenamefont
  {Plekhanov}, \citenamefont {Vasi{\'c}}, \citenamefont {Petrescu},
  \citenamefont {Nirwan}, \citenamefont {Roux}, \citenamefont {Hofstetter},\
  and\ \citenamefont {Le~Hur}}]{plekhanov2018emergent}%
  \BibitemOpen
  \bibfield  {author} {\bibinfo {author} {\bibfnamefont {K.}~\bibnamefont
  {Plekhanov}}, \bibinfo {author} {\bibfnamefont {I.}~\bibnamefont
  {Vasi{\'c}}}, \bibinfo {author} {\bibfnamefont {A.}~\bibnamefont {Petrescu}},
  \bibinfo {author} {\bibfnamefont {R.}~\bibnamefont {Nirwan}}, \bibinfo
  {author} {\bibfnamefont {G.}~\bibnamefont {Roux}}, \bibinfo {author}
  {\bibfnamefont {W.}~\bibnamefont {Hofstetter}}, \ and\ \bibinfo {author}
  {\bibfnamefont {K.}~\bibnamefont {Le~Hur}},\ }\href@noop {} {\bibfield
  {journal} {\bibinfo  {journal} {Phys. Rev. Lett.}\ }\textbf {\bibinfo
  {volume} {120}},\ \bibinfo {pages} {157201} (\bibinfo {year}
  {2018})}\BibitemShut {NoStop}%
\bibitem [{\citenamefont {Orignac}\ and\ \citenamefont
  {Giamarchi}(2001)}]{Orignac2000Meissner}%
  \BibitemOpen
  \bibfield  {author} {\bibinfo {author} {\bibfnamefont {E.}~\bibnamefont
  {Orignac}}\ and\ \bibinfo {author} {\bibfnamefont {T.}~\bibnamefont
  {Giamarchi}},\ }\href {\doibase 10.1103/PhysRevB.64.144515} {\bibfield
  {journal} {\bibinfo  {journal} {Phys. Rev. B}\ }\textbf {\bibinfo {volume}
  {64}},\ \bibinfo {pages} {144515} (\bibinfo {year} {2001})}\BibitemShut
  {NoStop}%
\bibitem [{\citenamefont {Boada}\ \emph {et~al.}(2012)\citenamefont {Boada},
  \citenamefont {Celi}, \citenamefont {Latorre},\ and\ \citenamefont
  {Lewenstein}}]{PhysRevLett.108.133001}%
  \BibitemOpen
  \bibfield  {author} {\bibinfo {author} {\bibfnamefont {O.}~\bibnamefont
  {Boada}}, \bibinfo {author} {\bibfnamefont {A.}~\bibnamefont {Celi}},
  \bibinfo {author} {\bibfnamefont {J.~I.}\ \bibnamefont {Latorre}}, \ and\
  \bibinfo {author} {\bibfnamefont {M.}~\bibnamefont {Lewenstein}},\ }\href
  {\doibase 10.1103/PhysRevLett.108.133001} {\bibfield  {journal} {\bibinfo
  {journal} {Phys. Rev. Lett.}\ }\textbf {\bibinfo {volume} {108}},\ \bibinfo
  {pages} {133001} (\bibinfo {year} {2012})}\BibitemShut {NoStop}%
\bibitem [{\citenamefont {Petrescu}\ and\ \citenamefont
  {Le~Hur}(2013)}]{PhysRevLett.111.150601}%
  \BibitemOpen
  \bibfield  {author} {\bibinfo {author} {\bibfnamefont {A.}~\bibnamefont
  {Petrescu}}\ and\ \bibinfo {author} {\bibfnamefont {K.}~\bibnamefont
  {Le~Hur}},\ }\href {\doibase 10.1103/PhysRevLett.111.150601} {\bibfield
  {journal} {\bibinfo  {journal} {Phys. Rev. Lett.}\ }\textbf {\bibinfo
  {volume} {111}},\ \bibinfo {pages} {150601} (\bibinfo {year}
  {2013})}\BibitemShut {NoStop}%
\bibitem [{\citenamefont {Celi}\ \emph {et~al.}(2014)\citenamefont {Celi},
  \citenamefont {Massignan}, \citenamefont {Ruseckas}, \citenamefont {Goldman},
  \citenamefont {Spielman}, \citenamefont {Juzeli\ifmmode~\bar{u}\else
  \={u}\fi{}nas},\ and\ \citenamefont {Lewenstein}}]{PhysRevLett.112.043001}%
  \BibitemOpen
  \bibfield  {author} {\bibinfo {author} {\bibfnamefont {A.}~\bibnamefont
  {Celi}}, \bibinfo {author} {\bibfnamefont {P.}~\bibnamefont {Massignan}},
  \bibinfo {author} {\bibfnamefont {J.}~\bibnamefont {Ruseckas}}, \bibinfo
  {author} {\bibfnamefont {N.}~\bibnamefont {Goldman}}, \bibinfo {author}
  {\bibfnamefont {I.~B.}\ \bibnamefont {Spielman}}, \bibinfo {author}
  {\bibfnamefont {G.}~\bibnamefont {Juzeli\ifmmode~\bar{u}\else
  \={u}\fi{}nas}}, \ and\ \bibinfo {author} {\bibfnamefont {M.}~\bibnamefont
  {Lewenstein}},\ }\href {\doibase 10.1103/PhysRevLett.112.043001} {\bibfield
  {journal} {\bibinfo  {journal} {Phys. Rev. Lett.}\ }\textbf {\bibinfo
  {volume} {112}},\ \bibinfo {pages} {043001} (\bibinfo {year}
  {2014})}\BibitemShut {NoStop}%
\bibitem [{\citenamefont {Petrescu}\ and\ \citenamefont
  {Le~Hur}(2015)}]{PhysRevB.91.054520}%
  \BibitemOpen
  \bibfield  {author} {\bibinfo {author} {\bibfnamefont {A.}~\bibnamefont
  {Petrescu}}\ and\ \bibinfo {author} {\bibfnamefont {K.}~\bibnamefont
  {Le~Hur}},\ }\href {\doibase 10.1103/PhysRevB.91.054520} {\bibfield
  {journal} {\bibinfo  {journal} {Phys. Rev. B}\ }\textbf {\bibinfo {volume}
  {91}},\ \bibinfo {pages} {054520} (\bibinfo {year} {2015})}\BibitemShut
  {NoStop}%
\bibitem [{\citenamefont {Tokuno}\ and\ \citenamefont
  {Georges}(2014)}]{Tokuno_2014}%
  \BibitemOpen
  \bibfield  {author} {\bibinfo {author} {\bibfnamefont {A.}~\bibnamefont
  {Tokuno}}\ and\ \bibinfo {author} {\bibfnamefont {A.}~\bibnamefont
  {Georges}},\ }\href {\doibase 10.1088/1367-2630/16/7/073005} {\bibfield
  {journal} {\bibinfo  {journal} {New Journal of Physics}\ }\textbf {\bibinfo
  {volume} {16}},\ \bibinfo {pages} {073005} (\bibinfo {year}
  {2014})}\BibitemShut {NoStop}%
\bibitem [{\citenamefont {Piraud}\ \emph {et~al.}(2015)\citenamefont {Piraud},
  \citenamefont {Heidrich-Meisner}, \citenamefont {McCulloch}, \citenamefont
  {Greschner}, \citenamefont {Vekua},\ and\ \citenamefont
  {Schollw\"ock}}]{PhysRevB.91.140406}%
  \BibitemOpen
  \bibfield  {author} {\bibinfo {author} {\bibfnamefont {M.}~\bibnamefont
  {Piraud}}, \bibinfo {author} {\bibfnamefont {F.}~\bibnamefont
  {Heidrich-Meisner}}, \bibinfo {author} {\bibfnamefont {I.~P.}\ \bibnamefont
  {McCulloch}}, \bibinfo {author} {\bibfnamefont {S.}~\bibnamefont
  {Greschner}}, \bibinfo {author} {\bibfnamefont {T.}~\bibnamefont {Vekua}}, \
  and\ \bibinfo {author} {\bibfnamefont {U.}~\bibnamefont {Schollw\"ock}},\
  }\href {\doibase 10.1103/PhysRevB.91.140406} {\bibfield  {journal} {\bibinfo
  {journal} {Phys. Rev. B}\ }\textbf {\bibinfo {volume} {91}},\ \bibinfo
  {pages} {140406} (\bibinfo {year} {2015})}\BibitemShut {NoStop}%
\bibitem [{\citenamefont {Greschner}\ \emph {et~al.}(2016)\citenamefont
  {Greschner}, \citenamefont {Piraud}, \citenamefont {Heidrich-Meisner},
  \citenamefont {McCulloch}, \citenamefont {Schollw\"ock},\ and\ \citenamefont
  {Vekua}}]{PhysRevA.94.063628}%
  \BibitemOpen
  \bibfield  {author} {\bibinfo {author} {\bibfnamefont {S.}~\bibnamefont
  {Greschner}}, \bibinfo {author} {\bibfnamefont {M.}~\bibnamefont {Piraud}},
  \bibinfo {author} {\bibfnamefont {F.}~\bibnamefont {Heidrich-Meisner}},
  \bibinfo {author} {\bibfnamefont {I.~P.}\ \bibnamefont {McCulloch}}, \bibinfo
  {author} {\bibfnamefont {U.}~\bibnamefont {Schollw\"ock}}, \ and\ \bibinfo
  {author} {\bibfnamefont {T.}~\bibnamefont {Vekua}},\ }\href {\doibase
  10.1103/PhysRevA.94.063628} {\bibfield  {journal} {\bibinfo  {journal} {Phys.
  Rev. A}\ }\textbf {\bibinfo {volume} {94}},\ \bibinfo {pages} {063628}
  (\bibinfo {year} {2016})}\BibitemShut {NoStop}%
\bibitem [{\citenamefont {Anisimovas}\ \emph {et~al.}(2016)\citenamefont
  {Anisimovas}, \citenamefont {Ra\ifmmode \check{c}\else
  \v{c}\fi{}i\ifmmode~\bar{u}\else \={u}\fi{}nas}, \citenamefont {Str\"ater},
  \citenamefont {Eckardt}, \citenamefont {Spielman},\ and\ \citenamefont
  {Juzeli\ifmmode~\bar{u}\else \={u}\fi{}nas}}]{Anisimovas2016Semi}%
  \BibitemOpen
  \bibfield  {author} {\bibinfo {author} {\bibfnamefont {E.}~\bibnamefont
  {Anisimovas}}, \bibinfo {author} {\bibfnamefont {M.}~\bibnamefont {Ra\ifmmode
  \check{c}\else \v{c}\fi{}i\ifmmode~\bar{u}\else \={u}\fi{}nas}}, \bibinfo
  {author} {\bibfnamefont {C.}~\bibnamefont {Str\"ater}}, \bibinfo {author}
  {\bibfnamefont {A.}~\bibnamefont {Eckardt}}, \bibinfo {author} {\bibfnamefont
  {I.~B.}\ \bibnamefont {Spielman}}, \ and\ \bibinfo {author} {\bibfnamefont
  {G.}~\bibnamefont {Juzeli\ifmmode~\bar{u}\else \={u}\fi{}nas}},\ }\href
  {\doibase 10.1103/PhysRevA.94.063632} {\bibfield  {journal} {\bibinfo
  {journal} {Phys. Rev. A}\ }\textbf {\bibinfo {volume} {94}},\ \bibinfo
  {pages} {063632} (\bibinfo {year} {2016})}\BibitemShut {NoStop}%
\bibitem [{\citenamefont {Calvanese~Strinati}\ \emph
  {et~al.}(2017)\citenamefont {Calvanese~Strinati}, \citenamefont {Cornfeld},
  \citenamefont {Rossini}, \citenamefont {Barbarino}, \citenamefont {Dalmonte},
  \citenamefont {Fazio}, \citenamefont {Sela},\ and\ \citenamefont
  {Mazza}}]{PhysRevX.7.021033}%
  \BibitemOpen
  \bibfield  {author} {\bibinfo {author} {\bibfnamefont {M.}~\bibnamefont
  {Calvanese~Strinati}}, \bibinfo {author} {\bibfnamefont {E.}~\bibnamefont
  {Cornfeld}}, \bibinfo {author} {\bibfnamefont {D.}~\bibnamefont {Rossini}},
  \bibinfo {author} {\bibfnamefont {S.}~\bibnamefont {Barbarino}}, \bibinfo
  {author} {\bibfnamefont {M.}~\bibnamefont {Dalmonte}}, \bibinfo {author}
  {\bibfnamefont {R.}~\bibnamefont {Fazio}}, \bibinfo {author} {\bibfnamefont
  {E.}~\bibnamefont {Sela}}, \ and\ \bibinfo {author} {\bibfnamefont
  {L.}~\bibnamefont {Mazza}},\ }\href {\doibase 10.1103/PhysRevX.7.021033}
  {\bibfield  {journal} {\bibinfo  {journal} {Phys. Rev. X}\ }\textbf {\bibinfo
  {volume} {7}},\ \bibinfo {pages} {021033} (\bibinfo {year}
  {2017})}\BibitemShut {NoStop}%
\bibitem [{\citenamefont {Price}\ \emph {et~al.}(2017)\citenamefont {Price},
  \citenamefont {Ozawa},\ and\ \citenamefont {Goldman}}]{PhysRevA.95.023607}%
  \BibitemOpen
  \bibfield  {author} {\bibinfo {author} {\bibfnamefont {H.~M.}\ \bibnamefont
  {Price}}, \bibinfo {author} {\bibfnamefont {T.}~\bibnamefont {Ozawa}}, \ and\
  \bibinfo {author} {\bibfnamefont {N.}~\bibnamefont {Goldman}},\ }\href
  {\doibase 10.1103/PhysRevA.95.023607} {\bibfield  {journal} {\bibinfo
  {journal} {Phys. Rev. A}\ }\textbf {\bibinfo {volume} {95}},\ \bibinfo
  {pages} {023607} (\bibinfo {year} {2017})}\BibitemShut {NoStop}%
\bibitem [{\citenamefont {Sundar}\ \emph {et~al.}(2018)\citenamefont {Sundar},
  \citenamefont {Gadway},\ and\ \citenamefont {Hazzard}}]{sundar2018synthetic}%
  \BibitemOpen
  \bibfield  {author} {\bibinfo {author} {\bibfnamefont {B.}~\bibnamefont
  {Sundar}}, \bibinfo {author} {\bibfnamefont {B.}~\bibnamefont {Gadway}}, \
  and\ \bibinfo {author} {\bibfnamefont {K.~R.}\ \bibnamefont {Hazzard}},\
  }\href@noop {} {\bibfield  {journal} {\bibinfo  {journal} {Scientific
  reports}\ }\textbf {\bibinfo {volume} {8}},\ \bibinfo {pages} {3422}
  (\bibinfo {year} {2018})}\BibitemShut {NoStop}%
\bibitem [{\citenamefont {L.~Barbiero}\ and\ \citenamefont
  {Goldman}(2019)}]{Barbiero_ladder_2019}%
  \BibitemOpen
  \bibfield  {author} {\bibinfo {author} {\bibfnamefont {S.~N.}\ \bibnamefont
  {L.~Barbiero}, \bibfnamefont {L.~Chomaz}}\ and\ \bibinfo {author}
  {\bibfnamefont {N.}~\bibnamefont {Goldman}},\ }\href@noop {} {\bibfield
  {journal} {\bibinfo  {journal} {arXiv preprint arXiv:1907.10555}\ } (\bibinfo
  {year} {2019})}\BibitemShut {NoStop}%
\bibitem [{\citenamefont {Li}\ \emph {et~al.}(2011)\citenamefont {Li},
  \citenamefont {Bakhtiari}, \citenamefont {He},\ and\ \citenamefont
  {Hofstetter}}]{Li2011}%
  \BibitemOpen
  \bibfield  {author} {\bibinfo {author} {\bibfnamefont {Y.-Q.}\ \bibnamefont
  {Li}}, \bibinfo {author} {\bibfnamefont {M.~R.}\ \bibnamefont {Bakhtiari}},
  \bibinfo {author} {\bibfnamefont {L.}~\bibnamefont {He}}, \ and\ \bibinfo
  {author} {\bibfnamefont {W.}~\bibnamefont {Hofstetter}},\ }\href {\doibase
  10.1103/PhysRevB.84.144411} {\bibfield  {journal} {\bibinfo  {journal} {Phys.
  Rev. B}\ }\textbf {\bibinfo {volume} {84}},\ \bibinfo {pages} {144411}
  (\bibinfo {year} {2011})}\BibitemShut {NoStop}%
\bibitem [{\citenamefont {Anders}\ \emph {et~al.}(2010)\citenamefont {Anders},
  \citenamefont {Gull}, \citenamefont {Pollet}, \citenamefont {Troyer},\ and\
  \citenamefont {Werner}}]{PhysRevLett.105.096402}%
  \BibitemOpen
  \bibfield  {author} {\bibinfo {author} {\bibfnamefont {P.}~\bibnamefont
  {Anders}}, \bibinfo {author} {\bibfnamefont {E.}~\bibnamefont {Gull}},
  \bibinfo {author} {\bibfnamefont {L.}~\bibnamefont {Pollet}}, \bibinfo
  {author} {\bibfnamefont {M.}~\bibnamefont {Troyer}}, \ and\ \bibinfo {author}
  {\bibfnamefont {P.}~\bibnamefont {Werner}},\ }\href {\doibase
  10.1103/PhysRevLett.105.096402} {\bibfield  {journal} {\bibinfo  {journal}
  {Phys. Rev. Lett.}\ }\textbf {\bibinfo {volume} {105}},\ \bibinfo {pages}
  {096402} (\bibinfo {year} {2010})}\BibitemShut {NoStop}%
\bibitem [{\citenamefont {Urban}\ \emph {et~al.}(2009)\citenamefont {Urban},
  \citenamefont {Johnson}, \citenamefont {Henage}, \citenamefont {Isenhower},
  \citenamefont {Yavuz}, \citenamefont {Walker},\ and\ \citenamefont
  {Saffman}}]{urban_observation_2009}%
  \BibitemOpen
  \bibfield  {author} {\bibinfo {author} {\bibfnamefont {E.}~\bibnamefont
  {Urban}}, \bibinfo {author} {\bibfnamefont {T.~A.}\ \bibnamefont {Johnson}},
  \bibinfo {author} {\bibfnamefont {T.}~\bibnamefont {Henage}}, \bibinfo
  {author} {\bibfnamefont {L.}~\bibnamefont {Isenhower}}, \bibinfo {author}
  {\bibfnamefont {D.~D.}\ \bibnamefont {Yavuz}}, \bibinfo {author}
  {\bibfnamefont {T.~G.}\ \bibnamefont {Walker}}, \ and\ \bibinfo {author}
  {\bibfnamefont {M.}~\bibnamefont {Saffman}},\ }\href {\doibase
  10.1038/nphys1178} {\bibfield  {journal} {\bibinfo  {journal} {Nat. Phys.}\
  }\textbf {\bibinfo {volume} {5}},\ \bibinfo {pages} {110} (\bibinfo {year}
  {2009})}\BibitemShut {NoStop}%
\bibitem [{\citenamefont {Gaetan}\ \emph {et~al.}(2009)\citenamefont {Gaetan},
  \citenamefont {Miroshnychenko}, \citenamefont {Wilk}, \citenamefont {Chotia},
  \citenamefont {Viteau}, \citenamefont {Comparat}, \citenamefont {Pillet},
  \citenamefont {Browaeys},\ and\ \citenamefont
  {Grangier}}]{gaetan_observation_2009}%
  \BibitemOpen
  \bibfield  {author} {\bibinfo {author} {\bibfnamefont {A.}~\bibnamefont
  {Gaetan}}, \bibinfo {author} {\bibfnamefont {Y.}~\bibnamefont
  {Miroshnychenko}}, \bibinfo {author} {\bibfnamefont {T.}~\bibnamefont
  {Wilk}}, \bibinfo {author} {\bibfnamefont {A.}~\bibnamefont {Chotia}},
  \bibinfo {author} {\bibfnamefont {M.}~\bibnamefont {Viteau}}, \bibinfo
  {author} {\bibfnamefont {D.}~\bibnamefont {Comparat}}, \bibinfo {author}
  {\bibfnamefont {P.}~\bibnamefont {Pillet}}, \bibinfo {author} {\bibfnamefont
  {A.}~\bibnamefont {Browaeys}}, \ and\ \bibinfo {author} {\bibfnamefont
  {P.}~\bibnamefont {Grangier}},\ }\href {\doibase 10.1038/nphys1183}
  {\bibfield  {journal} {\bibinfo  {journal} {Nat. Phys.}\ }\textbf {\bibinfo
  {volume} {5}},\ \bibinfo {pages} {115} (\bibinfo {year} {2009})}\BibitemShut
  {NoStop}%
\bibitem [{\citenamefont {Choi}\ and\ \citenamefont
  {Doniach}(1985)}]{PhysRevB.31.4516}%
  \BibitemOpen
  \bibfield  {author} {\bibinfo {author} {\bibfnamefont {M.~Y.}\ \bibnamefont
  {Choi}}\ and\ \bibinfo {author} {\bibfnamefont {S.}~\bibnamefont {Doniach}},\
  }\href {\doibase 10.1103/PhysRevB.31.4516} {\bibfield  {journal} {\bibinfo
  {journal} {Phys. Rev. B}\ }\textbf {\bibinfo {volume} {31}},\ \bibinfo
  {pages} {4516} (\bibinfo {year} {1985})}\BibitemShut {NoStop}%
\bibitem [{\citenamefont {Granato}\ and\ \citenamefont
  {Kosterlitz}(1986)}]{PhysRevB.33.4767}%
  \BibitemOpen
  \bibfield  {author} {\bibinfo {author} {\bibfnamefont {E.}~\bibnamefont
  {Granato}}\ and\ \bibinfo {author} {\bibfnamefont {J.~M.}\ \bibnamefont
  {Kosterlitz}},\ }\href {\doibase 10.1103/PhysRevB.33.4767} {\bibfield
  {journal} {\bibinfo  {journal} {Phys. Rev. B}\ }\textbf {\bibinfo {volume}
  {33}},\ \bibinfo {pages} {4767} (\bibinfo {year} {1986})}\BibitemShut
  {NoStop}%
\bibitem [{\citenamefont {O'Hern}\ and\ \citenamefont
  {Lubensky}(1998)}]{PhysRevLett.80.4345}%
  \BibitemOpen
  \bibfield  {author} {\bibinfo {author} {\bibfnamefont {C.~S.}\ \bibnamefont
  {O'Hern}}\ and\ \bibinfo {author} {\bibfnamefont {T.~C.}\ \bibnamefont
  {Lubensky}},\ }\href {\doibase 10.1103/PhysRevLett.80.4345} {\bibfield
  {journal} {\bibinfo  {journal} {Phys. Rev. Lett.}\ }\textbf {\bibinfo
  {volume} {80}},\ \bibinfo {pages} {4345} (\bibinfo {year}
  {1998})}\BibitemShut {NoStop}%
\bibitem [{\citenamefont {O'Hern}\ \emph {et~al.}(1999)\citenamefont {O'Hern},
  \citenamefont {Lubensky},\ and\ \citenamefont {Toner}}]{PhysRevLett.83.2745}%
  \BibitemOpen
  \bibfield  {author} {\bibinfo {author} {\bibfnamefont {C.~S.}\ \bibnamefont
  {O'Hern}}, \bibinfo {author} {\bibfnamefont {T.~C.}\ \bibnamefont
  {Lubensky}}, \ and\ \bibinfo {author} {\bibfnamefont {J.}~\bibnamefont
  {Toner}},\ }\href {\doibase 10.1103/PhysRevLett.83.2745} {\bibfield
  {journal} {\bibinfo  {journal} {Phys. Rev. Lett.}\ }\textbf {\bibinfo
  {volume} {83}},\ \bibinfo {pages} {2745} (\bibinfo {year}
  {1999})}\BibitemShut {NoStop}%
\bibitem [{\citenamefont {Pekker}\ \emph {et~al.}(2010)\citenamefont {Pekker},
  \citenamefont {Refael},\ and\ \citenamefont
  {Demler}}]{PhysRevLett.105.085302}%
  \BibitemOpen
  \bibfield  {author} {\bibinfo {author} {\bibfnamefont {D.}~\bibnamefont
  {Pekker}}, \bibinfo {author} {\bibfnamefont {G.}~\bibnamefont {Refael}}, \
  and\ \bibinfo {author} {\bibfnamefont {E.}~\bibnamefont {Demler}},\ }\href
  {\doibase 10.1103/PhysRevLett.105.085302} {\bibfield  {journal} {\bibinfo
  {journal} {Phys. Rev. Lett.}\ }\textbf {\bibinfo {volume} {105}},\ \bibinfo
  {pages} {085302} (\bibinfo {year} {2010})}\BibitemShut {NoStop}%
\bibitem [{\citenamefont {Niu}\ \emph {et~al.}(2018)\citenamefont {Niu},
  \citenamefont {Jin}, \citenamefont {Chen}, \citenamefont {Li},\ and\
  \citenamefont {Zhou}}]{Linxiao2018Observation}%
  \BibitemOpen
  \bibfield  {author} {\bibinfo {author} {\bibfnamefont {L.}~\bibnamefont
  {Niu}}, \bibinfo {author} {\bibfnamefont {S.}~\bibnamefont {Jin}}, \bibinfo
  {author} {\bibfnamefont {X.}~\bibnamefont {Chen}}, \bibinfo {author}
  {\bibfnamefont {X.}~\bibnamefont {Li}}, \ and\ \bibinfo {author}
  {\bibfnamefont {X.}~\bibnamefont {Zhou}},\ }\href@noop {} {\bibfield
  {journal} {\bibinfo  {journal} {Phys. Rev. Lett.}\ }\textbf {\bibinfo
  {volume} {121}} (\bibinfo {year} {2018})}\BibitemShut {NoStop}%
\bibitem [{\citenamefont {Chen}\ \emph {et~al.}(2018)\citenamefont {Chen},
  \citenamefont {Wang}, \citenamefont {Meng}, \citenamefont {Huang},
  \citenamefont {Cai}, \citenamefont {Wang}, \citenamefont {Zhu},\ and\
  \citenamefont {Zhang}}]{PhysRevLett.120.193601}%
  \BibitemOpen
  \bibfield  {author} {\bibinfo {author} {\bibfnamefont {L.}~\bibnamefont
  {Chen}}, \bibinfo {author} {\bibfnamefont {P.}~\bibnamefont {Wang}}, \bibinfo
  {author} {\bibfnamefont {Z.}~\bibnamefont {Meng}}, \bibinfo {author}
  {\bibfnamefont {L.}~\bibnamefont {Huang}}, \bibinfo {author} {\bibfnamefont
  {H.}~\bibnamefont {Cai}}, \bibinfo {author} {\bibfnamefont {D.-W.}\
  \bibnamefont {Wang}}, \bibinfo {author} {\bibfnamefont {S.-Y.}\ \bibnamefont
  {Zhu}}, \ and\ \bibinfo {author} {\bibfnamefont {J.}~\bibnamefont {Zhang}},\
  }\href {\doibase 10.1103/PhysRevLett.120.193601} {\bibfield  {journal}
  {\bibinfo  {journal} {Phys. Rev. Lett.}\ }\textbf {\bibinfo {volume} {120}},\
  \bibinfo {pages} {193601} (\bibinfo {year} {2018})}\BibitemShut {NoStop}%
\bibitem [{\citenamefont {Georges}\ \emph {et~al.}(1996)\citenamefont
  {Georges}, \citenamefont {Kotliar}, \citenamefont {Krauth},\ and\
  \citenamefont {Rozenberg}}]{Appen_georges96}%
  \BibitemOpen
  \bibfield  {author} {\bibinfo {author} {\bibfnamefont {A.}~\bibnamefont
  {Georges}}, \bibinfo {author} {\bibfnamefont {G.}~\bibnamefont {Kotliar}},
  \bibinfo {author} {\bibfnamefont {W.}~\bibnamefont {Krauth}}, \ and\ \bibinfo
  {author} {\bibfnamefont {M.~J.}\ \bibnamefont {Rozenberg}},\ }\href@noop {}
  {\bibfield  {journal} {\bibinfo  {journal} {Rev. Mod. Phys.}\ }\textbf
  {\bibinfo {volume} {68}},\ \bibinfo {pages} {13} (\bibinfo {year}
  {1996})}\BibitemShut {NoStop}%
\bibitem [{\citenamefont {Byczuk}\ and\ \citenamefont
  {Vollhardt}(2008)}]{Appen_Byczuk_2008}%
  \BibitemOpen
  \bibfield  {author} {\bibinfo {author} {\bibfnamefont {K.}~\bibnamefont
  {Byczuk}}\ and\ \bibinfo {author} {\bibfnamefont {D.}~\bibnamefont
  {Vollhardt}},\ }\href@noop {} {\bibfield  {journal} {\bibinfo  {journal}
  {Phys. Rev. B}\ }\textbf {\bibinfo {volume} {77}},\ \bibinfo {pages} {235106}
  (\bibinfo {year} {2008})}\BibitemShut {NoStop}%
\bibitem [{\citenamefont {Svidzinsky}\ and\ \citenamefont
  {Chang}(2008)}]{Svidzinsky_08}%
  \BibitemOpen
  \bibfield  {author} {\bibinfo {author} {\bibfnamefont {A.}~\bibnamefont
  {Svidzinsky}}\ and\ \bibinfo {author} {\bibfnamefont {J.-T.}\ \bibnamefont
  {Chang}},\ }\href {\doibase 10.1103/PhysRevA.77.043833} {\bibfield  {journal}
  {\bibinfo  {journal} {Phys. Rev. A}\ }\textbf {\bibinfo {volume} {77}},\
  \bibinfo {pages} {043833} (\bibinfo {year} {2008})}\BibitemShut {NoStop}%
\bibitem [{\citenamefont {Scully}\ and\ \citenamefont
  {Zubairy}(1997)}]{scully_quantum_1997}%
  \BibitemOpen
  \bibfield  {author} {\bibinfo {author} {\bibfnamefont {M.~O.}\ \bibnamefont
  {Scully}}\ and\ \bibinfo {author} {\bibfnamefont {M.~S.}\ \bibnamefont
  {Zubairy}},\ }\href@noop {} {\emph {\bibinfo {title} {Quantum {Optics}}}},\
  \bibinfo {edition} {1st}\ ed.\ (\bibinfo  {publisher} {Cambridge University
  Press},\ \bibinfo {year} {1997})\BibitemShut {NoStop}%
\bibitem [{\citenamefont {Daley}(2014)}]{daley_quantum_2014}%
  \BibitemOpen
  \bibfield  {author} {\bibinfo {author} {\bibfnamefont {A.~J.}\ \bibnamefont
  {Daley}},\ }\href {http://arxiv.org/abs/1405.6694} {\bibfield  {journal}
  {\bibinfo  {journal} {arXiv:1405.6694 [cond-mat, physics:quant-ph]}\ }
  (\bibinfo {year} {2014})}\BibitemShut {NoStop}%
\bibitem [{\citenamefont {Gadway}(2015)}]{gadway_atom-optics_2015}%
  \BibitemOpen
  \bibfield  {author} {\bibinfo {author} {\bibfnamefont {B.}~\bibnamefont
  {Gadway}},\ }\href {\doibase 10.1103/PhysRevA.92.043606} {\bibfield
  {journal} {\bibinfo  {journal} {Phys. Rev. A}\ }\textbf {\bibinfo {volume}
  {92}},\ \bibinfo {pages} {043606} (\bibinfo {year} {2015})}\BibitemShut
  {NoStop}%
\bibitem [{\citenamefont {An}\ \emph {et~al.}(2017)\citenamefont {An},
  \citenamefont {Meier},\ and\ \citenamefont {Gadway}}]{an2017direct}%
  \BibitemOpen
  \bibfield  {author} {\bibinfo {author} {\bibfnamefont {F.~A.}\ \bibnamefont
  {An}}, \bibinfo {author} {\bibfnamefont {E.~J.}\ \bibnamefont {Meier}}, \
  and\ \bibinfo {author} {\bibfnamefont {B.}~\bibnamefont {Gadway}},\
  }\href@noop {} {\bibfield  {journal} {\bibinfo  {journal} {Science advances}\
  }\textbf {\bibinfo {volume} {3}},\ \bibinfo {pages} {e1602685} (\bibinfo
  {year} {2017})}\BibitemShut {NoStop}%
\bibitem [{\citenamefont {An}\ \emph {et~al.}(2018{\natexlab{a}})\citenamefont
  {An}, \citenamefont {Meier}, \citenamefont {Ang'ong'a},\ and\ \citenamefont
  {Gadway}}]{an_correlated_2018}%
  \BibitemOpen
  \bibfield  {author} {\bibinfo {author} {\bibfnamefont {F.~A.}\ \bibnamefont
  {An}}, \bibinfo {author} {\bibfnamefont {E.~J.}\ \bibnamefont {Meier}},
  \bibinfo {author} {\bibfnamefont {J.}~\bibnamefont {Ang'ong'a}}, \ and\
  \bibinfo {author} {\bibfnamefont {B.}~\bibnamefont {Gadway}},\ }\href
  {\doibase 10.1103/PhysRevLett.120.040407} {\bibfield  {journal} {\bibinfo
  {journal} {Phys. Rev. Lett.}\ }\textbf {\bibinfo {volume} {120}},\ \bibinfo
  {pages} {040407} (\bibinfo {year} {2018}{\natexlab{a}})}\BibitemShut
  {NoStop}%
\bibitem [{\citenamefont {An}\ \emph {et~al.}(2018{\natexlab{b}})\citenamefont
  {An}, \citenamefont {Meier},\ and\ \citenamefont
  {Gadway}}]{an_engineering_2018}%
  \BibitemOpen
  \bibfield  {author} {\bibinfo {author} {\bibfnamefont {F.~A.}\ \bibnamefont
  {An}}, \bibinfo {author} {\bibfnamefont {E.~J.}\ \bibnamefont {Meier}}, \
  and\ \bibinfo {author} {\bibfnamefont {B.}~\bibnamefont {Gadway}},\ }\href
  {\doibase 10.1103/PhysRevX.8.031045} {\bibfield  {journal} {\bibinfo
  {journal} {Phys. Rev. X}\ }\textbf {\bibinfo {volume} {8}},\ \bibinfo {pages}
  {031045} (\bibinfo {year} {2018}{\natexlab{b}})}\BibitemShut {NoStop}%
\bibitem [{\citenamefont {Meier}\ \emph {et~al.}(2018)\citenamefont {Meier},
  \citenamefont {An}, \citenamefont {Dauphin}, \citenamefont {Maffei},
  \citenamefont {Massignan}, \citenamefont {Hughes},\ and\ \citenamefont
  {Gadway}}]{Meier929}%
  \BibitemOpen
  \bibfield  {author} {\bibinfo {author} {\bibfnamefont {E.~J.}\ \bibnamefont
  {Meier}}, \bibinfo {author} {\bibfnamefont {F.~A.}\ \bibnamefont {An}},
  \bibinfo {author} {\bibfnamefont {A.}~\bibnamefont {Dauphin}}, \bibinfo
  {author} {\bibfnamefont {M.}~\bibnamefont {Maffei}}, \bibinfo {author}
  {\bibfnamefont {P.}~\bibnamefont {Massignan}}, \bibinfo {author}
  {\bibfnamefont {T.~L.}\ \bibnamefont {Hughes}}, \ and\ \bibinfo {author}
  {\bibfnamefont {B.}~\bibnamefont {Gadway}},\ }\href {\doibase
  10.1126/science.aat3406} {\bibfield  {journal} {\bibinfo  {journal}
  {Science}\ }\textbf {\bibinfo {volume} {362}},\ \bibinfo {pages} {929}
  (\bibinfo {year} {2018})}\BibitemShut {NoStop}%
\bibitem [{\citenamefont {Scully}\ \emph {et~al.}(2006)\citenamefont {Scully},
  \citenamefont {Fry}, \citenamefont {Ooi},\ and\ \citenamefont
  {W\'odkiewicz}}]{scully_directed_2006}%
  \BibitemOpen
  \bibfield  {author} {\bibinfo {author} {\bibfnamefont {M.~O.}\ \bibnamefont
  {Scully}}, \bibinfo {author} {\bibfnamefont {E.~S.}\ \bibnamefont {Fry}},
  \bibinfo {author} {\bibfnamefont {C.~H.~R.}\ \bibnamefont {Ooi}}, \ and\
  \bibinfo {author} {\bibfnamefont {K.}~\bibnamefont {W\'odkiewicz}},\ }\href
  {\doibase 10.1103/PhysRevLett.96.010501} {\bibfield  {journal} {\bibinfo
  {journal} {Phys. Rev. Lett.}\ }\textbf {\bibinfo {volume} {96}},\ \bibinfo
  {pages} {010501} (\bibinfo {year} {2006})}\BibitemShut {NoStop}%
\bibitem [{\citenamefont {Fleischhauer}\ and\ \citenamefont
  {Lukin}(2000)}]{Fleischhauer_00}%
  \BibitemOpen
  \bibfield  {author} {\bibinfo {author} {\bibfnamefont {M.}~\bibnamefont
  {Fleischhauer}}\ and\ \bibinfo {author} {\bibfnamefont {M.~D.}\ \bibnamefont
  {Lukin}},\ }\href {\doibase 10.1103/PhysRevLett.84.5094} {\bibfield
  {journal} {\bibinfo  {journal} {Phys. Rev. Lett.}\ }\textbf {\bibinfo
  {volume} {84}},\ \bibinfo {pages} {5094} (\bibinfo {year}
  {2000})}\BibitemShut {NoStop}%
\bibitem [{\citenamefont {Wang}\ and\ \citenamefont
  {Scully}(2014)}]{wang2014heisenberg}%
  \BibitemOpen
  \bibfield  {author} {\bibinfo {author} {\bibfnamefont {D.-W.}\ \bibnamefont
  {Wang}}\ and\ \bibinfo {author} {\bibfnamefont {M.~O.}\ \bibnamefont
  {Scully}},\ }\href@noop {} {\bibfield  {journal} {\bibinfo  {journal}
  {Physical Review Letters}\ }\textbf {\bibinfo {volume} {113}},\ \bibinfo
  {pages} {083601} (\bibinfo {year} {2014})}\BibitemShut {NoStop}%
\end{thebibliography}%
\end{document}